\documentclass[onecolumn,showpacs,amsmath,amssymb]{revtex4}     
\usepackage{graphicx}  
\usepackage{bm}

\begin{document}  
  
\newcommand{\half}{\mbox{$\textstyle \frac{1}{2}$}}  
\newcommand{\ket}[1]{\left | \, #1 \right \rangle}  
\newcommand{\bra}[1]{\left \langle #1 \, \right |}  
\newcommand{\beq}{\begin{equation}}  
\newcommand{\eeq}{\end{equation}}  
\newcommand{\bea}{\begin{eqnarray}}  
\newcommand{\eea}{\end{eqnarray}}  
\newcommand{\req}[1]{Eq.\ (\ref{#1})}  
\newcommand{\gcc}{{\rm~g\,cm}^{-3}}  
\newcommand{\Compton}{\lambda\hspace{-.44em}\raisebox{.6ex}{\mbox{-$\!$-}}%
\raisebox{-.3ex}{}_{\hspace{-1pt}{\mbox{$_\mathrm{C}$}}}}  
\newcommand{\kB}{k_\mathrm{B}}  
\newcommand{\omc}{\omega_\mathrm{c}}  
\newcommand{\omg}{\omega_\mathrm{g}}  
\newcommand{\mel}{m_e}  
\newcommand{\xr}{x_{\rm r}}  
\newcommand{\EF}{\epsilon_{\rm F}}  
\newcommand{\Ne}{{\cal N}_B(\epsilon)}  
\newcommand{\Necl}{{\cal N}_0(\epsilon)}  
\newcommand{\dfde}{{\partial f^{(0)} \over\partial\epsilon}}  
\newcommand{\am}{a_\mathrm{m}}  
\newcommand{\dd}{{\rm\,d}}  
\newcommand{\vB}{\bm{B}}  
\newcommand{\dotZ}{\mbox{$\dot{\mbox{Z}}$}}  
\newcommand{\msun}{\mbox{$M_\odot$}}

\def\la{\;\raise0.3ex\hbox{$<$\kern-0.75em\raise-1.1ex\hbox{$\sim$}}\;}  
\def\ga{\;\raise0.3ex\hbox{$>$\kern-0.75em\raise-1.1ex\hbox{$\sim$}}\;}  
\def\lesssim{\;\raise0.3ex\hbox{$<$\kern-0.75em\raise-1.1ex\hbox{$\sim$}}\;}  
\def\gtrsim{\;\raise0.3ex\hbox{$>$\kern-0.75em\raise-1.1ex\hbox{$\sim$}}\;}  
  
\title{Nuclear fusion in dense matter:  
Reaction rate and carbon burning}  
  
\author{L. R. Gasques$^1$, A. V. Afanasjev$^1$, E. F. Aguilera$^2$,  
M. Beard$^1$, L. C. Chamon$^3$, P. Ring$^4$, M. Wiescher$^1$, and D. G.  
Yakovlev$^5$}  
\affiliation{$^1$ Department of Physics $\&$ The Joint Institute  
for Nuclear Astrophysics, University of Notre Dame,  Notre Dame,  
IN 46556 USA. \\  
$^2$ Departamento del Accelerador, Instituto Nacional de Investigaciones  
Nucleares, A.P. 18-1027, C.P. 11801, Destrito Federal, Mexico. \\  
$^3$ Departamento de F\'{\i}sica Nuclear, Instituto de F\'{\i}sica da  
Universidade de S\~ao Paulo, Caixa Postal 66318, 05315-970, S\~ao Paulo, SP,  
Brazil. \\  
$^4$ Physik-Department, Technische Universitat M\"{u}nchen, D-85747,  
Garching, Germany. \\  
$^5$ Ioffe Physical Technical Institute, Poliekhnicheskaya 26, 194021  
St.-Petersburg, Russia.}  
  
\date{\today}  
\begin{abstract}  
In this paper we analyze the nuclear fusion rate between equal  
nuclei for all five different nuclear burning regimes in dense matter  
(two thermonuclear regimes, two pycnonuclear ones, and the  
intermediate regime). The rate is determined by Coulomb barrier  
penetration in dense environments and by the astrophysical  
$S$-factor at low energies. We evaluate previous studies of the  
Coulomb barrier problem and propose a simple phenomenological  
formula for the reaction rate which covers all cases. The  
parameters of this formula can be varied, taking into account  
current theoretical uncertainties in the reaction rate. The  
results are illustrated for the example of the $^{12}$C+$^{12}$C  
fusion reaction. This reaction is very important for the  
understanding of nuclear burning in evolved stars, in exploding  
white dwarfs producing type Ia supernovae, and in accreting  
neutron stars. The $S$-factor at stellar energies depends on a  
reliable fit and extrapolation of the experimental data. We  
calculate the energy dependence of the $S$-factor using a recently  
developed parameter-free model for the nuclear interaction, taking  
into account the effects of the Pauli nonlocality.  For  
illustration, we analyze the efficiency of carbon burning in a  
wide range of densities and temperatures of stellar matter with  
the emphasis on carbon ignition at densities $\rho \gtrsim 10^9$  
g~cm$^{-3}$.  
\end{abstract}  
  
  
\maketitle  
  
\section{Introduction}  
\label{introduct}  
  
\indent We will study nuclear fusion rates of identical nuclei in  
dense stellar matter. This problem is of utmost importance for  
understanding the structure and evolution of stars of various  
types. Despite the efforts of many authors the theoretical  
reaction rates are still rather uncertain, especially at high  
densities. The uncertainties have two aspects. The first one is  
related to {\it nuclear physics} and is concerned with the proper  
treatment of {\it nuclear} interaction transitions (conveniently  
described in terms of the astrophysical factor $S$). The other  
issue is associated with aspects of {\it plasma physics} and  
concerns the proper description of Coulomb barrier penetration in  
a high density many-body system. We will analyze both aspects  
and illustrate the results taking the carbon fusion reaction as  
an example.  
  
Considerable experimental effort has been spent on the study of  
low energy fusion reactions such as $^{12}$C+$^{12}$C, to  
investigate the impact on the nucleosynthesis, energy  
production and time scale of late stellar evolution. Nevertheless,  
it has been difficult to develop a global and reliable reaction  
formalism to extrapolate the energy dependence of the fusion cross  
section into the stellar energy range. The overall energy  
dependence of the cross section is determined by the  
Coulomb-barrier tunnel probability. One goal of the present work  
is to apply the S\~ao Paulo potential model to provide a general 
description of the stellar fusion processes. This model does not contain 
any free parameter and represents a powerful tool to predict average low 
energy cross sections for a wide range of fusion reactions, as long as the 
density distribution of the nuclei involved in the reaction can be determined.
In this context we also seek to introduce a phenomenological formalism for  
a generalized reaction rate to describe all the regimes of  
nuclear burning in a one-component plasma ion system. In this  
paper we want to demonstrate the applicability of the method on  
the specific example of $^{12}$C+$^{12}$C, to evaluate the  
reliability and uncertainty range of the proposed formalism  
through the comparison with the available low energy data. In a  
subsequent publication we want to extend the model to  
multi-ion systems with the aim of simulating a broad range of heavy-ion  
nucleosynthesis scenarios from thermonuclear burning in hot  
stellar plasma, to pycnonuclear burning in high density crystalline  
stellar matter.  
  
Carbon burning represents the third phase of stellar evolution for  
massive stars ($M \gtrsim 8M_\odot$); it follows helium burning  
that converts He-fuel to $^{12}$C via the triple $\alpha$ process.  
Carbon burning represents the first stage during stellar evolution  
determined by heavy-ion fusion processes (e.g., Ref.\  
\cite{Wallerstein}). The most important reaction during the carbon  
burning phase is the $^{12}$C+$^{12}$C fusion \cite{Barnes};  
additional processes can be $^{12}$C+$^{16}$O and  
$^{16}$O+$^{16}$O, depending on the $^{12}$C/$^{16}$O abundance  
ratio which is determined by the $^{12}$C($\alpha,\gamma$)$^{16}$O  
reaction rate \cite{WW93,WHR03}. The most important reaction  
branches are $^{12}$C($^{12}$C,$\alpha$)$^{20}$Ne ($Q$=4.617 MeV)  
and $^{12}$C($^{12}$C,$p$)$^{23}$Na ($Q$=2.241 MeV). Carbon  
burning in evolved massive stars takes place at typical densities  
$\rho\sim 10^5$ g~cm$^{-3}$ and temperatures $T\sim (6-8)\times  
10^8$~K.  
  
Carbon burning is also crucial for type Ia supernovae. These  
supernova explosions are driven by carbon ignition in cores of  
accreting massive CO white dwarfs \cite{SNI-1}. The burning  
process proceeds from the carbon ignition region near the center  
of a white dwarf by detonation or deflagration through the entire  
white dwarf body. The ignition conditions and time scale are  
defined by the $^{12}$C+$^{12}$C reaction rate, typically, at  
$T\sim (1.5-7) \times 10^8$~K and $\rho \sim (2-5) \times  
10^9$~g~cm$^{-3}$ \cite{SNI,bhw04}. 
Depending on the $^{16}$O abundance, other fusion reactions may also
contribute. For these  high densities the reaction cross sections are affected 
by strong plasma screening, which reduces the repulsive Coulomb barrier  
between interacting $^{12}$C or $^{16}$O nuclei (e.g., Refs.\  
\cite{Salp,CLL02}; also see Section \ref{fusion}).  
  
Explosive carbon burning in the crust of accreting neutron stars  
has recently been proposed as a possible trigger and energy source  
for superbursts \cite{CB01,SB02,cn04}. In this scenario small  
amounts of carbon (3\%--10\%), which have survived in the  
preceding rp-process phase during the thermonuclear runaway,  
ignite after the rp-process ashes are compressed by accretion to a  
density $\rho \sim  1.3 \times 10^9$ g~cm$^{-3}$. The ignition of  
a carbon flash requires an initial temperature $T\gtrsim 10^9$~K,  
triggering a photodisintegration runaway of the rp-process ashes  
after a critical temperature  $T \sim 2 \times 10^9$ K is reached  
\cite{SBC03}. For these scenarios, carbon burning proceeds in the  
{\it thermonuclear} regime with strong plasma screening (see  
Section \ref{fusion} for details).  
  
At high densities and/or low temperatures the thermonuclear reaction rate 
formalism is insufficient since the fusion process is mainly driven by the 
high density conditions in  
stellar matter (Sections \ref{fusion} and \ref{carbonburn}). This  
is particularly important for nuclear fusion in the deeper layers  
of the crust of an accreting neutron star \cite{hz}. At  
sufficiently high $\rho$ and low $T$ nuclei form a crystalline  
lattice. Neighboring nuclei may penetrate the Coulomb barrier and fuse owing 
to zero-point vibrations in their  
lattice sites. In this {\it pycnonuclear} burning regime the  
reaction rate depends mainly on the density and is nearly  
independent of temperature (e.g. \cite{svh69,Schramm}).  
Pycnonuclear burning regimes may not be limited to carbon induced  
fusion reactions only, but may be driven by a broad range of fusion  
reactions between stable and neutron rich isotopes \cite{hz}.  
  
In the following Section \ref{sfactor}, we discuss the theory of  
fusion cross sections and calculate the astrophysical $S$-factor  
for carbon burning in the framework of a generalized  
parameter-free potential model. In Section \ref{fusion} we study  
the Coulomb barrier problem for identical nuclei, and propose an  
expression for the reaction rate which describes all the regimes  
of nuclear burning in a dense one-component plasma of atomic  
nuclei. In Section \ref{carbonburn} we analyze, for illustration,  
the main features of $^{12}$C burning from high temperature gaseous or liquid 
plasma to high density crystalline matter. We summarize and conclude in  
Section \ref{conclusions}.  
  
\section{Fusion Cross Section and Astrophysical $S$-factor}  
\label{sfactor}  
  
\indent Nuclear reactions are possible after colliding nuclei  
tunnel through the Coulomb barrier. Recently, a parameter-free  
model for the real part of the nuclear interaction (S\~ao Paulo  
potential) based on nonlocal quantum effects was developed  
\cite{ref6,ref7,ref8,Toward}. In previous work \cite{Fusion}, this  
model was applied to the study of fusion processes using the  
barrier penetration (BP) formalism for about 2500 cross section  
data, corresponding to approximately 165 different systems. Within  
the nonlocal model, the bare interaction $V_{N}(r,E)$ is connected  
with the folding potential $V_{F}(r)$,  
\begin{equation}  
    V_{N}(r,E)=V_F(r) \; e^{- 4 {\rm v}^2/c^2} \;,  
\label{nuc_pot}  
\end{equation}  
where $c$ is the speed of light, $E$ is the particle collision energy (in the  
center-of-mass reference frame), $\mbox{v}$ is the local relative velocity  
of the two nuclei 1 and 2,  
\begin{equation}  
    \mbox{v}^2(r,E) = \frac{2}{\mu} \,  
    \left[ E-V_C(r)-V_{N}(r,E) \right] \;,  
\label{speed}  
\end{equation}  
$V_C(r)$ is the Coulomb potential,  
$\mu = A_1 A_2 m_{\rm u}/ (A_1 + A_2)$ is the reduced mass,  
and $m_{\rm u}$ is the atomic mass unit.  The folding potential depends on 
the matter densities of the nuclei involved in the collision:  
\begin{equation}  
    V_F(R) = \int \rho_1( {\bm r}_1) \; \rho_2({\bm r}_2) \;  
    V_{0} \; \delta({\bm R}-{\bm r}_{1}+{\bm r}_{2}) \; d{\bm r}_1 ,  
\label{fold}  
\end{equation}  
with $V_0 = - 456$ MeV fm$^3$. The use of the matter densities and  
delta function in Eq.~(\ref{fold}) corresponds to the zero-range  
approach for the folding potential, which is equivalent  
\cite{Toward} to the more usual procedure of using the M3Y  
effective nucleon-nucleon interaction with the nucleon densities  
of the nuclei. The advantage in adopting the S\~ao Paulo potential  
to describe the fusion cross section relies on the fact that no  
additional parameter is necessary once the density distribution of  
the participating nuclei has been determined. The model is  
therefore a good choice for a generalized treatment of low energy  
heavy-ion fusion reactions.  
  
There are several ways to determine the nuclear density  
distribution  
\cite{Toward,Afanasjev,Afanasjev1,Gonzalez,NL2,NL3,DD-ME1,DD-ME2}.  
Density Functional Theories (DFT) provide for example a successful  
description of many nuclear ground state properties, in 
particular,  
of charge distributions in the experimentally known region. Since  
these theories are universal in the sense that their parameter  
sets are carefully adjusted and valid all over the periodic table,  
one can expect that they also yield reliable predictions for  
nuclei far from stability. Non-relativistic density functionals,  
such as the Skyrme or Gogny functional, have been widely used in  
the literature. In recent years, relativistic density functionals  
have played an increasingly important role since they provide a  
fully consistent description of the spin-orbit splitting. This is  
of greatest importance for nuclei far from stability. The  
spin-orbit splitting determines the shell structure, the most  
basic ingredient in any microscopic theory of finite nuclei. In  
fact, the results obtained with relativistic functionals are in  
very good agreement with experimental data, throughout the  
periodic table, despite having a smaller number of adjustable  
parameters in comparison with the non-relativistic case. Best  
known is the relativistic Hartree-Bogoliubov theory  
\cite{Afanasjev,Afanasjev1,Gonzalez}, which includes pairing  
correlations with finite range pairing forces. It provides a  
unified description of mean-field and pairing correlations in  
nuclei.  

These functionals contain a strong density dependence, either  
through non-linear coupling terms between the meson fields  
(e.g., in the Lagrangians with the parameter sets NL2  
\cite{NL2} and NL3 \cite{NL3}), or by using an explicit density  
dependence for the meson-nucleon vertices (e.g., in the  
parameter sets DD-ME1 \cite{DD-ME1} and DD-ME2 \cite{DD-ME2}).  
  
\begin{figure}[tbh]  
\begin{center}  
\includegraphics[width=13.0cm,angle=-90]{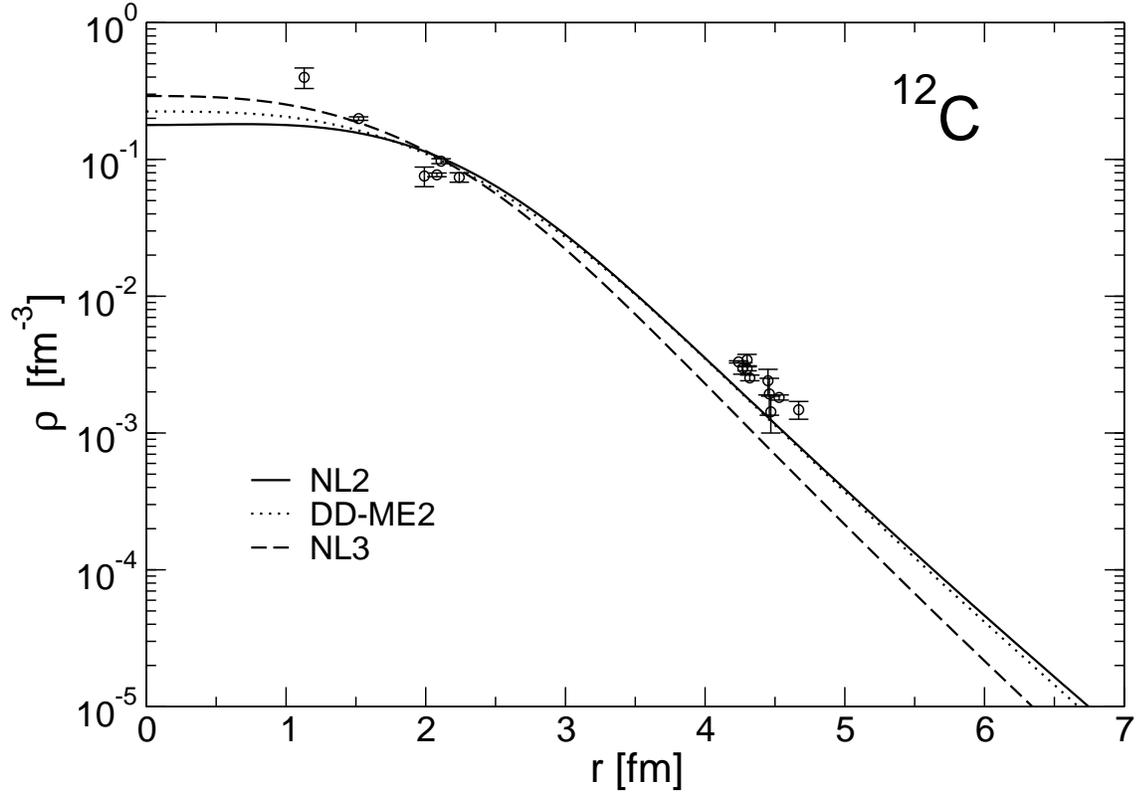}  
\caption{  
Self-consistent densities for the ground state of $^{12}$C calculated with  
different parameterizations of the RMF Lagrangian. The densities obtained 
with the DD-ME1 and DD-ME2 interactions are very similar. The experimental 
data are taken from Ref.\ \cite{c12-dens}.}  
\label{fig-dens}  
\end{center}  
\end{figure}

In the present paper, we consider only spherical nuclear shapes.  
Pairing correlations are in principle included, but they vanish  
for the $^{12}$C nucleus. 
In Fig.~\ref{fig-dens} we compare the  
calculated densities with experimental data \cite{c12-dens}. The  
RHB calculations are in good agreement with surface properties  
best described by the NL2, DD-ME1, and DD-ME2 effective  
interactions.  
  
To apply the BP model for calculating fusion cross sections one  
needs the effective potential defined as a sum of the Coulomb,  
nuclear and centrifugal components:  
\begin{equation}  
V_{\rm eff} (r,E) = V_{C} (r) + V_{N} (r,E) + \frac{\ell(\ell+1)\hbar^2}  
{2 \mu r^2}.  
\label{eff_pot}  
\end{equation}  
Following the BP model one can associate the fusion cross section with the  
particle flux transmitted through the barrier,  
\begin{equation}  
  \sigma_{ij} (E) = \frac{\pi} {k^2}\; \sum_{\ell=0}^{\ell_{cr}} (2\ell+1)\;  
T_{\ell}.  
\label{sigma}  
\end{equation}  
It is important to point out that the sum in Eq.\ (\ref{sigma}) is  
performed up to a maximum $\ell$ wave ($\ell_{cr})$, which  
corresponds to the greatest value of angular momentum that  
produces a pocket (and a barrier) in the corresponding effective  
potential, Eq.\ (\ref{eff_pot}). For $\ell$-waves with effective  
barrier heights $V_{B\ell}<E$, the shape of the effective  
potential can be approximated by a parabola with curvature defined  
as  
\begin{equation}  
    \hbar \omega_{\ell} =  
     \left|\frac{\hbar^2} {\mu}\; \frac{d^2V_{\rm eff}} {dr^2}\right|_  
    {R_{B\ell}}^{1/2},  
\label{hw1}  
\end{equation}  
where $R_{B\ell}$ is the barrier radius. In such cases, the  
transmission coefficients have been obtained through the  
Hill-Wheeler formula \cite{Hill}:  
\begin{equation}  
    T_{\ell} = \left\{1+\exp\left[\frac{2\pi(V_{B\ell}-E)}  
    {\hbar \omega_{\ell}}  
    \right]\right\}^{-1}.  
\label{hill}  
\end{equation}  
On the other hand, for $\ell$-waves with $V_{B\ell}>E$,  
instead of the Hill-Wheeler formula, we employ a more appropriate  
heuristic treatment based on a WKB approximation \cite{Schiff}:  
\begin{equation}  
   T_\ell = [1+\exp(S_{\ell})]^{-1},  
   \label{wkb}  
\end{equation}  
\begin{equation}  
  S_\ell = \int_{r_{1}}^{r_{2}} \sqrt{\frac{8 \mu} {\hbar^2}\, [V_{\rm eff}  
   (r,E)-E]} \; dr,  
\label{swkb}  
\end{equation}  
where $r_{1}$ and $r_{2}$ are the classical turning points. At low  
energies, the WKB method gives values for the transmission  
coefficients which are quite different from those of the  
Hill-Wheeler formula. In this case, we define the barrier  
curvature by connecting Eqs.\ (6) and (7):  
\begin{equation}  
   \hbar \omega_{\ell} = \frac{2\pi(V_{B\ell}-E)} {S_\ell}.  
\label{hw2}  
\end{equation}  

The overall results provided by the BP model are in very good agreement  
with the fusion data for energies above the s-wave barrier height.  
For light systems $(\mu \leq 8 \, m_{\rm u})$ the model also shows  
very good agreement with fusion data at sub-barrier energies  
\cite{Fusion}. Therefore, the use of the BP model in calculating  
the fusion cross section at energies of astrophysical interest for  
the $^{12}$C+$^{12}$C system is entirely justified.  
  
Historically, reaction cross sections $\sigma(E)$ at very low  
energies, typical for astrophysical conditions, have been  
expressed in terms of the astrophysical $S$-factor (e.g., Ref.\  
\cite{FCZII}),  
\begin{equation}  
   S(E) = \sigma(E) \, E \; e^{2\pi\eta} \ ,  
\label{sf}  
\end{equation}  
where $\eta=(Z_1 Z_2 e^2/\hbar) \sqrt{\mu /(2E)}$  
is the usual Gamow parameter.  
  
\begin{figure}[tbh]  
\begin{center}  
\includegraphics[width=12.0cm,angle=-90]{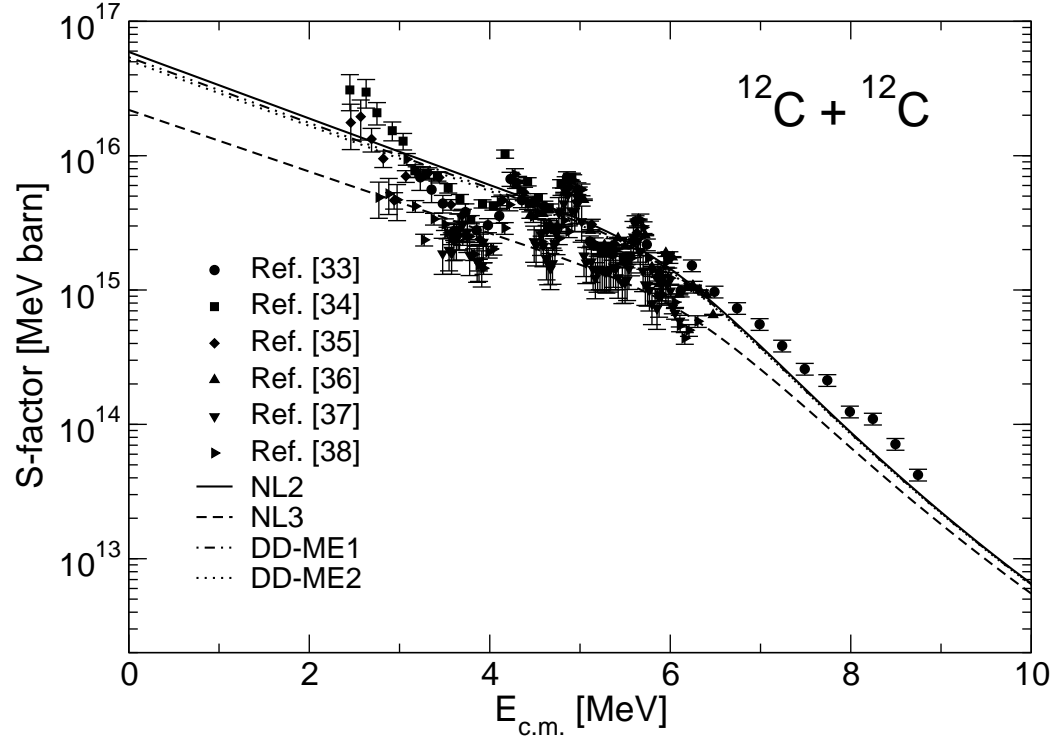}  
\caption{  
Astrophysical factor $S(E)$ as a function of the center-of-mass energy $E$,  
derived from experimentally measured cross sections. Lines show theoretical  
results obtained within the barrier penetration model for the different model  
density distributions (see text for details).  
Various symbols present experimental results.}  
\label{fig1}  
\end{center}  
\end{figure}

Considerable efforts have been made over the last decades to  
measure the $^{12}$C+$^{12}$C fusion cross section at very low  
energies \cite{Patterson,Mazarakis,High,Eli,Kettner,Becker}. The  
experimentally determined $S$-factors are shown in Fig.~\ref{fig1}. 
For reaction rate calculations the experimental  
$S$-factor needs to be extrapolated towards the stellar energy  
range, the Gamow window, which depends sensitively on the  
temperature and density conditions of the stellar environment. The  
typical range of energy $E$ for thermonuclear
carbon burning, in the center-of-mass reference, varies from 1 to 4 MeV.
For pycnonuclear carbon burning in the neutron star crusts
the energies can be as low as 10 keV.
Large discrepancies between the different 
experimental results at low energies 
complicate a reliable extrapolation of $S(E)$ towards such low $E$. In 
addition, the $S$-factor shows pronounced resonant structures, presumably  
resulting from quasimolecular doorway states. Theoretical  
calculations of $S(E)$ using the effective interactions NL2, NL3,  
DD-ME1, DD-ME2 agree reasonably well, within a factor $\sim$ 3.5  
in the limit $E$ $\rightarrow$ 0 (Fig. \ref{fig1}). Furthermore,  
the resonant behavior of the data cannot be described with the BP  
calculations because the effects of nuclear structure were  
neglected. However, an average description of the data (neglecting  
resonant oscillations) for the sub-barrier region ($E$ $\lesssim$  
6.0 MeV) is reproduced satisfactorily. Such a description of  
an average $S$-factor is quite sufficient since the reaction rate  
formalism relies on the average $S$-factor behavior over the  
entire Gamow range.  
  
In this context it is important to emphasize that the main purpose  
of this paper is not to investigate the oscillations in the  
$^{12}$C+$^{12}$C fusion excitation function. In order to  
reproduce the resonances we could, for example, use the concept of  
internal and barrier waves based on a semiclassical description  
\cite{Oh87} or adopt the R-matrix formalism (e.g., Michaud and  
Vogt \cite{Mi72}). However, neither theoretical approach 
would allow us to extrapolate with confidence the fusion cross section to the  
energy region of astrophysical interest.  
  
In order to calculate the carbon burning rate, we use the values  
of $S(E)$ obtained on the basis of the well established NL2  
effective interaction. As one can see from Fig. \ref{fig-dens},  
the $^{12}$C density distribution obtained using the parameter set  
NL2 can describe satisfactorily the surface properties, which is  
the most important region for the fusion process at low energies.  
The values of $S(E)$ calculated at $E \leq 19.8$ MeV can be fitted  
by an analytic expression  
\begin{equation}  
   S(E)=5.15 \times 10^{16}\,
   \exp \left\{- 0.428\,E - {3\,E^{0.308} \over
   1+{\rm e}^{0.613\,(8-E)}} \right\}~~{\rm MeV~barn},  
\label{fit}  
\end{equation}  
where the center-of-mass energy $E$ is expressed in MeV. The  
formal maximum fit error, 16\%, occurs at $E=5.8$ MeV. However, let  
us bear in mind that the values of $S(E)$ provided by the NL2  
model and given by Eq.\ (\ref{fit}) are actually uncertain within  
a factor of $\sim 3.5$.  
  
\begin{figure}[tbh]  
\begin{center}  
\includegraphics[width=12.0cm,angle=-90]{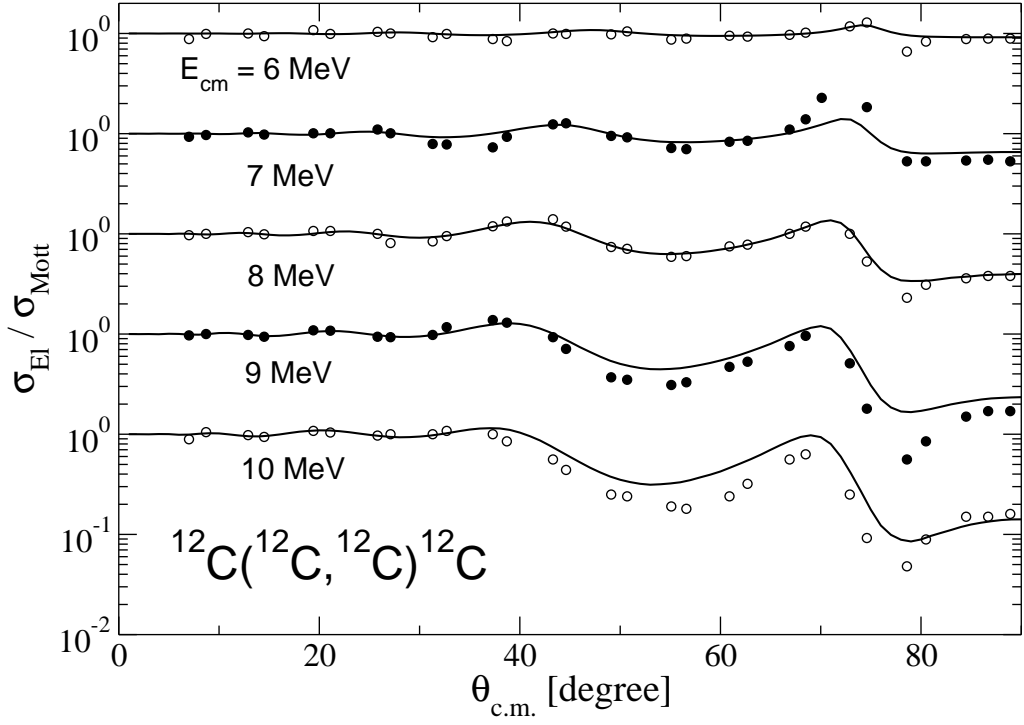}  
\caption{Elastic scattering angular distributions for the $^{12}$C+$^{12}$C 
system at energies around and slightly above the Coulomb barrier 
\cite{Treu80}. The lines are the results of an optical model calculation 
assuming the S\~ao Paulo potential to describe the real part of the nuclear 
interaction, combined with a simple model to describe the imaginary part of 
the optical potential (see details in the text).}  
\label{elas}  
\end{center}  
\end{figure}

With the aim of investigating the validity of our assumption for the real part 
of the nuclear interaction, we performed an optical model (OM) analysis of the 
$^{12}$C+$^{12}$C elastic scattering data at energies around and slightly 
above the Coulomb barrier \cite{Treu80}. We defined the imaginary part of the 
optical potential, which accounts for the nuclear absorption process, as
\begin{equation}
W(r,E) = N_i \, V_{N}(r,E),
\end{equation}
where $V_{N}(r,E)$ is described by Eq.~(\ref{nuc_pot}) and $N_i$ = 0.78 was 
determined by adjusting thirty elastic scattering angular distributions 
corresponding to seven different heavy-ion systems and measured in a very 
wide energy range \cite{Al03}. 
Figure~\ref{elas} illustrates a comparison 
between our OM analysis and five elastic scattering angular distribution data 
of the $^{12}$C+$^{12}$C system. As one can note, it is possible to obtain a 
reasonable description of the data by adopting the S\~ao Paulo potential 
to account for the real part of the nuclear interaction, combined 
with a simple model to describe the imaginary part of the optical potential. 
This means that both elastic scattering and fusion processes can be 
described by the same real part of the nuclear interaction, which has been 
well accounted by the S\~ao Paulo potential, Eq.~(\ref{nuc_pot}). 
As discussed in Ref.~\cite{Al03}, details on the absorption part of the 
interaction are not very important for describing the elastic scattering data, 
which allows us to get reasonable estimates for the $^{12}$C+$^{12}$C 
system.   

Further experiments at lower energies are necessary to confirm the  
validity of the predicted $^{12}$C+$^{12}$C $S$-factor and its  
impact on the reaction rate. However, the $S$-factor is not the only  
uncertainty for a reliable description of the $^{12}$C+$^{12}$C  
fusion process in stellar matter. The reaction rates are  
also uncertain because of the problems in calculating the  
probability of Coulomb barrier penetration in a dense many-body  
environment. We shall discuss these problems in Section \ref{fusion}  
and show that the associated uncertainties are higher than the current  
uncertainties  
in the values of $S(E)$.  
  
\section{Nuclear fusion rate in dense matter}  
\label{fusion}  
  
\subsection{Physical conditions and reaction regimes}  
\label{regimes}  
  
\indent In the following we will turn our attention to the plasma  
physics aspects of nuclear burning in dense matter. We will focus  
on the formalism of fusion reactions between identical nuclei  
$(A,Z)+(A,Z)$ in the wide domain of temperatures $T$ and densities  
$\rho$, characteristic for the range of stellar environments  
outlined above.  
  
\begin{figure}[tbh]  
\begin{center}  
\includegraphics[width=12.0cm]{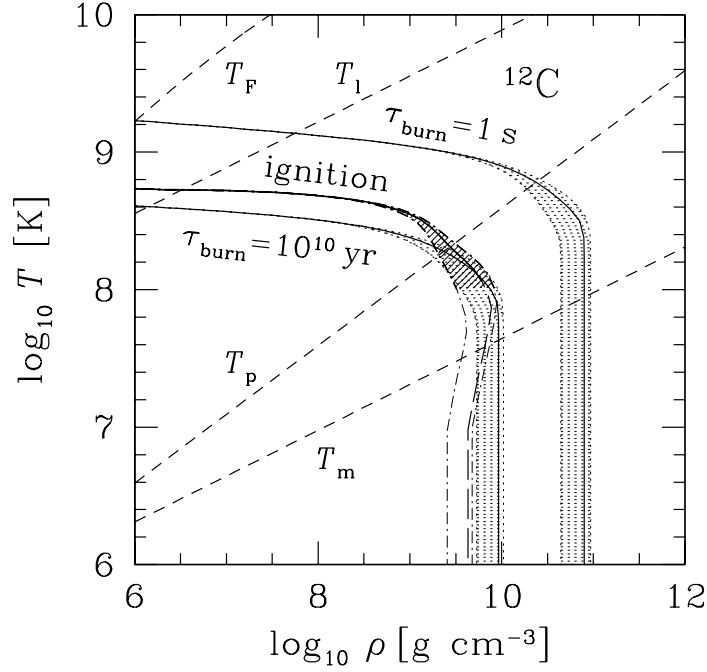}  
\caption{  
Temperature-density diagram for carbon matter.  
Short-dashed lines show the electron degeneracy temperature  
$T_{\rm F}$, the temperature $T_l$ of the appearance of ion  
liquid, the melting temperature  $T_{\rm m}$ of ion  
crystal, and the ion plasma temperature $T_p$.  
Solid lines correspond to the carbon burning times  
$\tau_{\rm burn}=1$~s and  $\tau_{\rm burn}=10^{10}$ yr,  
and to carbon ignition; they are calculated using the most  
reliable model of carbon burning (Section \ref{together}).  
Hatched strips show theoretical uncertainties of these lines  
(limited by the minimum and maximum reaction rate models).  
The long-dashed line exhibits the unreliable part of  
the ignition curve; nearby thin dashed-and-dot lines  
(to the right and left) indicate its assumed uncertainties.}  
\label{fig:diag}  
\end{center}  
\end{figure}

As an example, we consider the $^{12}$C+$^{12}$C reaction in  
stellar matter at conditions 
displayed in the $\rho-T$ phase  
diagram in Fig.\ \ref{fig:diag}. Under these conditions, carbon is  
fully ionized (either by electron pressure and/or by high  
temperature) and immersed in an almost uniform electron  
background. 
The electrons are typically
strongly degenerate; their  
degeneracy temperature $T_{\rm F}$ is shown in the figure.  
  
The state of ions (nuclei) is determined by the Coulomb coupling  
parameter $\Gamma = Z^2e^2/(aT)$, where $a=[3/(4 \pi n_i)]^{1/3}$  
is the ion-sphere radius and $n_i$ is the number density of ions;  
the Boltzmann constant is set $k_{\rm B}\equiv 1$. If $\Gamma \la  
1$ (which happens at $T \ga T_l =Z^2e^2/a$, see Fig.\  
\ref{fig:diag}), the ions constitute a Boltzmann gas, while at  
higher $\Gamma$ they constitute a strongly coupled Coulomb liquid.  
The gas transforms smoothly into the liquid, without any phase  
transition. At small $T$ (large $\Gamma$) the liquid can solidify.  
In the density range displayed in Fig.\ \ref{fig:diag}, the  
solidification occurs at $T=T_{\rm m}=Z^2e^2/a\Gamma_{\rm m}$,  
where $\Gamma_{\rm m}=175$ (e.g., De Witt et al.\  
\cite{dewittetal03}). The important measure of quantum effects in  
ion motion is provided by the ion plasma frequency  
$\omega_p=\sqrt{ 4 \pi Z^2 e^2 n_i /m}$ or the associated ion  
plasma temperature $T_p= \hbar \omega_p$ ($m$ being the ion mass).  
As a rule, the quantum effects are strongly pronounced at $T$  
below $T_p$.  
  
Figure \ref{fig:diag} shows that the ion  
system can have  very different properties, depending on $T$ and $\rho$. As a  
result, there are five qualitatively different regimes of nuclear burning in  
dense matter (Salpeter and Van Horn \cite{svh69}). These are (1)  
the classical thermonuclear regime; (2) the thermonuclear regime  
with strong plasma screening; (3) the thermo-pycnonuclear regime;  
(4) the thermally enhanced pycnonuclear regime; and (5) the  
zero-temperature pycnonuclear regime. The regimes differ mainly in  
the character of the Coulomb barrier penetration of reacting  
nuclei. The penetration can be greatly complicated by Coulomb  
fields of ions which surround the reacting nuclei. These fields  
are fluctuating and random (e.g., Alastuey and Jancovici  
\cite{aj78}).  
  
A strict solution of the barrier penetration problem should imply  
the calculation of the tunneling probability in a random  
potential, with subsequent averaging over an ensemble of random  
potentials. This program has not been fully realized so far. The  
exact theory should take into account a range of effects which can  
be subdivided (somewhat conventionally) into classical and quantum  
ones. The classical effects are associated with classical motion  
of plasma ions and with related structure of Coulomb plasma fields  
(including spatial and temporal variability of these fields). The  
quantum effects manifest themselves in ion motion (e.g.,  
zero-point ion vibrations), quantum ``widths'' of ion trajectories  
during Coulomb barrier penetration, and quantum statistics of  
reacting nuclei. The effects of quantum statistics are usually  
small due to the obvious reason that quantum tunneling lengths are  
typically much larger than nuclear radii. The smallness of these  
effects has been confirmed by Ogata \cite{ogata97} in  
path-integral Monte Carlo (PIMC) simulations.  
  
The reaction rates in the classical thermonuclear regime are well  
known (e.g., Fowler, Caughlan, and Zimmerman \cite{FCZII}); they  
have been tested very successfully by the theory and observations  
of the evolution of normal stars. This theory will be only shortly  
reviewed in the following section. The reaction rates in other  
regimes have been calculated by a number of authors in different  
approximations. In the following we summarize the main results  
published after the seminal paper by Salpeter and Van Horn  
\cite{svh69} (see that paper for references to earlier works). Let  
us stress that the reaction rate is a rapidly varying function of  
plasma parameters. In the most important density-temperature  
domain it varies over tens orders of magnitude (Section  
\ref{carbonburn}). In this situation, a very precise calculation  
of the reaction rate is very difficult but not required for many  
applications.  
  
\subsection{Classical thermonuclear reaction rate}  
\label{thermo}  
  
The classical thermonuclear regime takes place at sufficiently  
high $T$ and low $\rho$ so that the ions constitute a Boltzmann  
gas ($T \gg T_l$, Fig.\ \ref{fig:diag}). The tunnel probability  
(penetrability) through the Coulomb barrier depends on the energy  
of the interacting ions; the main contribution to the reaction  
rate comes from ion collisions with energies approximately equal  
to the Gamow peak energy $E_{\rm pk}$ (that is much higher than  
$T$). This regime is typical for all nuclear burning stages in ``normal''  
stars (from the main sequence to 
pre-supernovae).  
  
The thermonuclear reaction rate is expressed by  
\begin{equation}  
    R_{\rm th}= {n_i^2 \over 2} \, 4 \,  
    \sqrt{ 2 E_{\rm pk} \over 3 \mu} \,  
    { S(E_{\rm pk}) \over T } \, \exp(-\tau),  
\label{therm}  
\end{equation}  
where $E_{\rm pk}=T \tau /3$ is the Gamow peak energy and  
\begin{equation}  
  \tau= \left( 27 \pi^2 \mu Z_1^2 Z_2^2 e^4 \over 2 T \hbar^2 \right)^{1/3}  
   =  \left( 27 \pi^2 m Z^4 e^4 \over 4 T \hbar^2 \right)^{1/3}  
\label{tau}  
\end{equation}  
is the parameter which characterizes the penetrability $\sim  
\exp(-\tau$). The parameter $\tau$ can be rewritten as  
\begin{equation}  
    \tau=3\,(\pi/2)^{2/3}(E_{\rm a}/T)^{1/3}, \quad  
    E_{\rm a} \equiv m Z^4 e^4/\hbar^2.  
\label{tau1}  
\end{equation}  

Now the reaction rate can be presented as  
\begin{equation}  
    R_{\rm th}= { n_i^2 \over 2} \, S(E_{\rm pk}) \,  
     { \hbar \over m Z^2e^2 }\, P_{\rm th} \, F_{\rm th},  
\label{Rth}  
\end{equation}  
where $ \hbar /(m Z^2 e^2)$ is a convenient dimensional  
factor, $F_{\rm th}$ is the exponential  
function, and $P_{\rm th}$ is the pre-exponent:  
\begin{equation}  
     F_{\rm th}=\exp(-\tau), \quad  
     P_{\rm th}= { 8 \pi^{1/3} \over \sqrt{3}\, 2^{1/3}} \,  
       \left(E_{\rm a} \over T \right)^{2/3}.  
\label{Exth}  
\end{equation}  
The classical thermonuclear reaction rate decreases exponentially  
with decreasing $T$.  
  
\subsection{Thermonuclear burning with strong plasma screening}  
\label{thermoscreen}  
  
The thermonuclear regime with strong plasma screening operates in  
a colder and denser plasma ($T_p \la T \la T_l$), where ions  
constitute a strongly coupled classical Coulomb system (liquid or  
solid). The majority of ions in such a system are confined in deep  
Coulomb potential wells ($Z^2e^2/a \ga T$). The main contribution  
into the reaction rate comes from a small amount of higher-energy,  
unbound ions with $E \approx E_{\rm pk}\gg T$ from the tail of the  
Boltzmann distribution. The plasma screening effects are produced  
by surrounding plasma ions and simplify close approaches of the  
reacting nuclei, required for a successful Coulomb tunneling. This  
enhances the reaction rate with respect to that given by Eqs.\  
(\ref{Rth}) and (\ref{Exth}).  
  
The enhancement has been studied by a number of authors, beginning  
with Salpeter \cite{Salp}; it can reach many orders of magnitude.  
Calculations show that the equations given in Section \ref{thermo}  
remain valid in this regime, but the penetrability function  
$F_{\rm th}$ has to be corrected for the screening effects:  
\begin{equation}  
     F_{\rm th}=F_{\rm sc}\cdot\exp(-\tau)\, \quad F_{\rm sc}=\exp(h),  
\label{Exscr}  
\end{equation}  
where $F_{\rm sc}$ is the enhancement factor,  
and $h$ is a function of plasma parameters.  
  
Plasma screening effects are usually modeled by introducing a  
mean-force plasma potential $H(r)$. In this approximation, the  
reacting nuclei move in a potential $W(r)=Z^2e^2/r-H(r)$. The  
mean-force  
plasma potential $H(r)$ is static and spherically symmetric. It  
cannot take into account dynamical variations of plasma  
microfields and their instantaneous spatial structures in the  
course of an individual tunneling event. In the mean-force  
approximation, the function $h$ consists of two parts, $h=h_0+  
h_1$, where the leading term $h_0=H(0)/T$ ($\gg |h_1|$) is  
calculated assuming a constant plasma potential $H(r)=H(0)$ during  
the quantum tunneling, while $h_1$ is a correction due to a weak  
variation of $H(r)$ along the tunneling path. Note that according  
to simple estimates (e.g., Ref.\ \cite{ys89}) typical tunneling  
lengths of reacting ions in the thermonuclear regime (where $T  
\gtrsim T_p$) are considerably smaller than the ion sphere radius  
$a$, and typical tunneling times are much smaller than the plasma  
oscillation period $\sim \omega_p^{-1}$. This justifies the assumption  
of almost constant and static plasma potential during a tunneling  
event.  
  
The mean-force plasma potential $H(r)$ for a  
classical strongly coupled system of ions (liquid or solid)  
can be determined using  
classical Monte Carlo (MC) sampling (e.g., DeWitt et al.\  
\cite{dgc73}). MC sampling gives the static radial-pair  
distribution function of ions $g(r)=\exp(-W(r)/T)$ which enables  
one to find $H(r)$. In this way one can accurately determine  
$g(r)$ and $H(r)$ at not too small $r$ (typically, at $r \gtrsim  
a$), because of poor MC statistics of close ion separations. The  
potential $H(r)$ at small $r$, required for a tunneling problem,  
is obtained by extrapolating MC values of $H(r)$ to $r \to 0$; the  
extrapolation procedure is a delicate subject and may be ambiguous  
(as discussed, e.g., by Rosenfeld \cite{rosenfeld96}).  
  
It is only $H(0)$ which  
is required for finding $h_0$. For a classical ion  
system, $H(0)$ can be determined by $H(0)=\Delta {\cal F}$, where  
$\Delta {\cal F}$ is a difference of 
Coulomb free energies (for a  
given system and for a system 
with two nuclei merging into one compound nucleus;  
e.g., DeWitt et al.\ \cite{dgc73}). In this approximation, the  
enhancement factor of the nuclear reaction becomes a thermodynamic  
quantity and acquires a Boltzmann form, $\exp (h_0)= \exp( \Delta  
{\cal F}/T)$, showing that plasma screening increases the  
probability of close separations  
(and subsequent quantum tunneling); $h_0$ becomes the function of  
one argument $\Gamma$. Assuming a linear mixing rule in a  
multi-component strongly coupled ion system, Jancovici  
\cite{jancovici77} obtained $h_0=2f_0(\Gamma)-f_0(2^{5/3}\Gamma)$,  
where $f_0(\Gamma)$ is a Coulomb free energy of one ion in a  
one-component plasma of ions (in units of $T$). In a Coulomb  
liquid at $\Gamma \gtrsim 1$ the linear mixing rule is highly  
accurate (DeWitt and Slattery \cite{ds03}); the function  
$f_0(\Gamma)$ is now determined from MC sampling with very high  
accuracy (e.g., Ref.\ \cite{pc00,ds03}). In this way the function  
$h_0(\Gamma)$ has been calculated  
in many papers (e.g., Refs.\ \cite{jancovici77,ys89,rosenfeld96,ds99}),  
and the results are in very good agreement.  
Let us present the analytical approximation of $h_0(\Gamma)$  
which follows from the recent MC results  
of DeWitt and Slattery \cite{ds99} for a Coulomb liquid at  
$1 \leq \Gamma \leq  170$:  
\begin{equation}  
    h_0=1.0563 \, \Gamma + 1.0208\, \Gamma^{0.3231}  
      -0.2748 \, \ln \Gamma -1.0843.  
\label{accfit}  
\end{equation}  
However, this accurate expression is inconvenient for further use, and we  
propose another fit  
\begin{equation}  
     h_0=C_{\rm sc}\, \Gamma^{3/2}/  
    [(C_{\rm sc}/\sqrt{3})^4+\Gamma^2]^{1/4},  
\label{scrfit}  
\end{equation}  
where $C_{\rm sc}=1.0754$. It approximates ${\rm e}^{h_0}$  
with the maximum error of $\sim$40\% at $\Gamma=170$, quite  
sufficient for our purpose.  
There may be still some uncertainty of the reaction  
rate associated with the choice of $C_{\rm sc}$ but it seems  
to be not higher than the uncertainty in the $S$-factor  
(Section \ref{sfactor}).  
Our fit function in Eq.\  
(\ref{scrfit}) is chosen in such a way to reproduce also the well  
known expression $h_0 \to \sqrt{3}\,\Gamma^{3/2}$ derived by  
Salpeter \cite{Salp} for the classical thermonuclear regime  
($\Gamma \ll 1$), where  
$h_0 \ll 1$ and the plasma screening is weak. In the  
Coulomb liquid, at $\Gamma \gtrsim 1$, we have actually the linear  
function $h_0=C_{\rm sc} \Gamma$. Such a function was obtained by  
Salpeter \cite{Salp} using a simple model of ion spheres (with  
slightly lower coefficient, $C_{\rm sc}^{\rm Salp} = 1.057$).  
  
Some authors  
calculated $h_0$ and the associated enhancement factor  
${\rm e}^{h_0}$ by  
extrapolating MC $H(r)$ to $r \to 0$ (as discussed above).  
In particular, Ogata et al.\ \cite{oii91,oiv93} employed this formalism  
to study the enhancement of nuclear reactions  
in one-component and two-component  
strongly coupled ion liquids.  
The enhancement factor ${\rm e}^{h_0}$ for a one-component  
ion liquid, calculated in these papers (e.g., Eq.\ (6)  
in Ref.\ \cite{oii91}), is systematically higher  
than the factor given by Eq.\ (\ref{accfit}) or (\ref{scrfit}).  
The difference reaches a factor of approximately 40 for $\Gamma \sim 170$.  
Because the enhancement factor itself  
becomes as high as ${\rm e}^{h_0}\sim 10^{74}$ at $\Gamma \sim 170$,  
such a difference is insignificant for many applications.  
As shown by Rosenfeld \cite{rosenfeld96},  
the difference comes from the problems  
of extrapolation of $H(r)$ to $r \to 0$  
in Refs.\ \cite{oii91,oiv93}.  
The function  
$h_0$ was also calculated by Ogata \cite{ogata97} using direct  
PIMC method. His result (his Eq.\ (19)) is in much better  
agreement with Eq.\ (\ref{accfit}). The maximum difference of  
${\rm e}^{h_0}$ reaches only a factor of approximately 6 at $\Gamma=170$.  
Recently new PIMC calculations have been performed  
by Pollock and Militzer \cite{pm04} but the authors  
have not calculated directly $h_0(\Gamma)$.  
  
Let us emphasize that the enhancement factor  
${\rm e}^{h_0}$, derived in a constant  
mean-force plasma potential $H(0)$, is invariant  
with respect to the order of the mean-force averaging  
and the tunneling  
probability calculation. One can consider a real (random)  
plasma potential, constant over a tunneling path  
in an individual tunneling event. Calculating  
the tunneling probability and averaging  
over an ensemble of realizations of plasma potentials,  
one comes (e.g., Ref.\ \cite{ys89}) to the same expression for $h_0$  
as given by the mean-force potential.  
  
In addition to ${\rm e}^{h_0}$, the enhancement factor $F_{\rm  
sc}$ in Eq.\ (\ref{Exscr}) contains a smaller factor ${\rm  
e}^{h_1}$, associated with variations of the plasma potential  
along the tunneling path. Numerous calculations of $h_1$ have commonly  
employed the mean-force potential $H(r)$. The results are sensitive  
to the behavior of $H(r)$ at small $r$ (where this behavior is not  
very certain). For example, Jancovici \cite{jancovici77} got  
$h_1=-(5/32)\,\Gamma\,(3 \Gamma /\tau)^2$. Note that for the  
thermonuclear burning ($T \gtrsim T_p$, Section \ref{regimes}),  
the ratio $3\Gamma / \tau  
\approx r_{\rm t}/a \sim (T/T_p)^{2/3}$ can be regarded as a  
small parameter ($r_{\rm t}$ being the tunneling  
length). It is possible that the mean-force approximation is  
too crude for calculating $h_1$. For that reason, we will not  
specify $h_1$ in this section. Our final expression for the  
reaction rate will include $h_1$, but phenomenologically, when we  
combine reaction rates in all regimes (Section \ref{together}).

\subsection{Zero-temperature pycnonuclear fusion}  
\label{pycnozero}  
  
The zero-temperature pycnonuclear regime operates in a cold and  
dense matter ($T$ well below $T_p$) in a strongly coupled quantum  
system of nuclei. In this regime the Coulomb barrier is penetrated  
owing to zero-point vibrations of neighboring nuclei which occupy  
their ground states in a strongly coupled system. One usually  
considers pycnonuclear reactions in a crystalline lattice of  
nuclei but they are also possible in a quantum liquid. The main  
contribution into the reaction rate comes from pairs of nuclei  
which are most closely spaced. The reaction rate is  
temperature-independent but increases exponentially with  
increasing density as we will discuss in the following.  
  
Pycnonuclear reaction rates between identical nuclei in  
crystalline lattice have been calculated  
by many authors using different  approximations.  
In analogy with Eq.\ (\ref{Rth}), the resulting reaction rates  
 can be written as  
\begin{equation}  
    R_{\rm pyc}= { n_i^2 \over 2} \, S(E_{\rm pk}) \,  
     { \hbar \over m Z^2e^2 }\, P_{\rm pyc} \, F_{\rm pyc},  
\label{Rpyc}  
\end{equation}  
where $F_{\rm pyc}$ and $P_{\rm pyc}$ depend on the  
density and have the form  
\begin{equation}  
     F_{\rm pyc}=\exp \left(- C_{\rm exp} / \sqrt{\lambda} \right), \quad  
     P_{\rm pyc}= 8 \,C_{\rm pyc}\,11.515/\lambda^{C_{\rm pl}}.  
\label{Expyc}  
\end{equation}  
The dimensionless parameters $C_{\rm exp}$, $C_{\rm pl}$ and  
$C_{\rm pyc}$ are  
model dependent (see below). The dimensionless  
parameter $\lambda$ is expressed in terms of the mass fraction  
$X_i$ contained in atomic nuclei (in a one-component ion plasma  
under study) and the mass density $\rho$ of the medium  
\begin{equation}  
  \lambda= {\hbar^2 \over m Z^2 e^2 }\,  
  \left( n_i \over 2 \right)^{1/3}=  
  { 1 \over AZ^2 } \,  
  \left( {1 \over A} \, { \rho \, X_i \over 1.3574 \times 10^{11}~  
  {\rm g~cm}^{-3} } \right)^{1/3}.  
\label{lambda}  
\end{equation}  
For densities $\rho$ lower than the neutron drip density ($\sim 4  
\times 10^{11}$ g~cm$^{-3}$; e.g., Ref.\ \cite{st83}), one can set  
$X_i=1$, while for higher $\rho$ one has $X_i<1$ because of the  
presence of free (dripped) neutrons.  
  
The reaction rate can be expressed numerically as  
\begin{equation}  
    R_{\rm pyc}= \rho \, X_i \, AZ^4 \, S(E_{\rm pk}) \,  
    C_{\rm pyc}\,  
    10^{46} \, \lambda^{3-C_{\rm pl}}\,  
    \exp\left(-C_{\rm exp}/\sqrt{\lambda}\right)  
    ~~~{\rm s}^{-1}~{\rm cm}^{-3},  
\label{pyc-numer}  
\end{equation}  
where $\rho$ is in g~cm$^{-3}$ and $S(E_{\rm pk})$ is in  
MeV~barn. The typical energy of the interacting nuclei is  
$E_{\rm pk} \sim \hbar \omega_p$.  
  
\begin{figure}[tbh]  
\begin{center}  
\includegraphics[width=12.0cm]{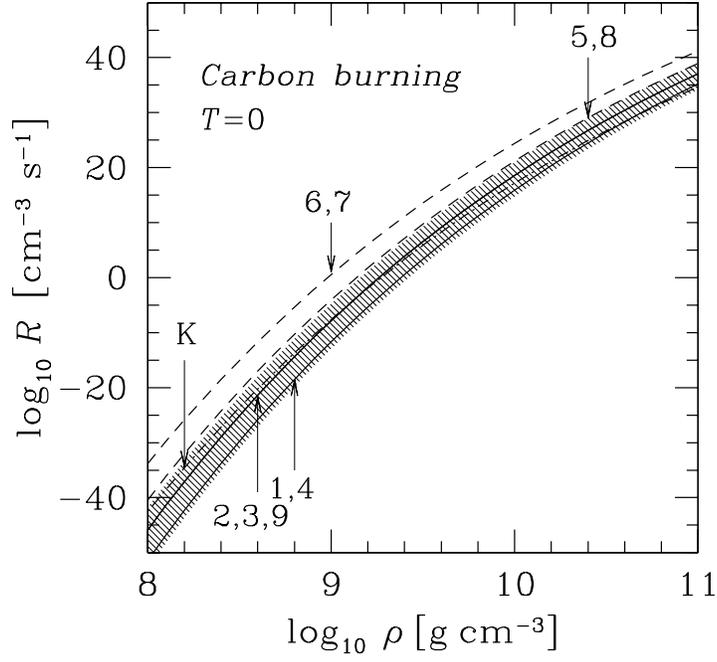}  
\caption{Rate of pycnonuclear carbon burning at $T=0$ as a function  
of density for the different theoretical models  
(from Table \ref{tab:pyc}). Solid and dashed  
lines refer to the burning in bcc and fcc crystals, respectively;  
the dash-and-dot line `K' is the model by Kitamura \cite{kitamura00}  
(for bcc crystal).  
Hatched strip shows assumed uncertainties of the reaction  
rates for bcc crystals (limited by models 10 and 11 from Table  
\ref{tab:pyc}).}  
\label{fig:pyc}  
\end{center}  
\end{figure}

Table \ref{tab:pyc} lists the values of $C_{\rm exp}$, $C_{\rm pl}$ and  
$C_{\rm pyc}$ reported in the literature for two models (1 and 2) of Coulomb  
barrier penetration by Salpeter and Van Horn \cite{svh69}, for  
six models (3--8) by Schramm and Koonin  
\cite{Schramm}, and for one model (9) by Ogata, Iyetomi and  
Ichimaru \cite{oii91}. The corresponding carbon burning rates are  
plotted as a function of density in Fig.\ \ref{fig:pyc}.  
In this figure (as well as in Figs.\ \ref{fig:diag}  
and \ref{fig:therm}) we use the astrophysical  
factors given by the fit expression (\ref{fit}).  
Actually, the $S$-factors are uncertain within one  
order of magnitude (Section \ref{sfactor}) but we ignore  
these  
uncertainties (because they seem to be much lower than those  
associated with the Coulomb barrier problem).  
  
All the authors cited above
have treated quantum tunneling by fixing the  
center-of-mass of reacting nuclei in its equilibrium position.  
All models, except for models 5--8, 
focus  
on nuclear reactions in the body-centered cubic (bcc) lattice of  
atomic nuclei. This lattice is thought to be preferable over other  
lattices, particularly, over the face-centered cubic (fcc)  
lattice. The main reason is that the bcc lattice is more tightly  
bound in the approximation of a rigid electron background.  
However, the difference in binding energies of bcc and fcc  
lattices is small (see, e.g., Ref.\ \cite{Schramm}), and a finite  
polarizability of the electron background complicates the problem  
\cite{baiko}. Therefore, one cannot exclude that  
the lattice type is fcc.  
  
Salpeter and Van Horn \cite{svh69} calculated the quantum  
tunneling probability of interacting nuclei in a bcc Coulomb  
lattice using the three-dimensional WKB approximation (most adequate  
for the given problem). The authors employed two models, {\it  
static} and {\it relaxed} lattice (models 1 and 2 in Table  
\ref{tab:pyc}), to account for the lattice response to the motion  
of tunneling nuclei. The static lattice model assumes that  
surrounding nuclei remain in their original lattice sites during  
the tunneling process. The relaxed lattice model assumes that the  
surrounding nuclei are  
promptly rearranged into new equilibrium positions  
in response to the motion of the reacting nuclei.  
Simple estimates show that the  
actual tunneling is dynamical (neither static not relaxed). Thus,  
the static-lattice and relaxed-lattice models impose constraints  
on the actual reaction rate. In Ref.\ \cite{svh69} the screening  
potential for the relaxed-lattice model was calculated  
approximately; the energy difference between the initial and fused  
states was evaluated by subtracting the energies of the  
corresponding Wigner-Seitz (WS) spheres. The relaxed lattice  
simplifies Coulomb tunneling and increases the reaction rate with  
respect to the static lattice (cf.\ curves 1 and 2 in Fig.\  
\ref{fig:pyc}).  
  
Schramm and Koonin \cite{Schramm} applied this treatment to the  
bcc and fcc, static  
and relaxed lattices in the same WKB approximation. They calculated the  
screening potential for the relaxed-lattice model with improved  
accuracy (model 3 of Table \ref{tab:pyc} for the bcc lattice and  
model 7 for fcc). For comparison, they also used the screening  
potential for the relaxed-lattice obtained in the WS approximation  
(as in Ref.\ \cite{svh69}). Unfortunately, they calculated the  
tunneling probability neglecting the correction ${\rm e}^K$ for to  
the ``curvature of trajectories'' of reacting ions.  This is the  
main reason for the formal disagreement between the results of  
Salpeter and Van Horn \cite{svh69} and Schramm and Koonin  
\cite{Schramm} for the static-lattice and relaxed-lattice-WS  
models (1 and 2) of bcc crystals. The inclusion of the curvature correction  
should reduce the constant $C_{\rm pyc}$ in Eq.\ (\ref{pyc-numer})  
and the reaction rates calculated in Ref.\ \cite{Schramm}.  
Fortunately, this correction can be extracted by comparing Eq.\ (38)  
of Ref.\ \cite{svh69} with Eq.\ (31) of Ref.\ \cite{Schramm}. In  
this way we get ${\rm e}^K=0.067$ for the static bcc lattice, and  
${\rm e}^K=0.050$ for the relaxed-WS bcc lattice. After  
introducing this correction into the coefficients $C_{\rm pyc}$,  
obtained formally from the results of Schramm and Koonin, these  
coefficients become identical to those given by Salpeter and Van  
Horn. Thus, Schramm and Koonin actually exactly reproduce models 1 and  
2 of Salpeter and Van Horn.  
The curvature correction for the models 3--8 of Schramm and  
Koonin have not been determined. We expect it to be ${\rm e}^K \approx  
0.050$ for model 3 and ${\rm e}^K=0.067$ for model 4 (bcc  
crystals), and we introduced such corrections in Table \ref{tab:pyc}.  
We introduced, somewhat arbitrarily, the correction ${\rm  
e}^K=0.05$ in all fcc lattice models 5--8.  
  
The two versions of the screening potential for the relaxed  
lattice (WS and more accurate) almost coincide. Accordingly,  
models 2 and 3 yield almost the same reaction rates for the bcc  
lattice, while models 6 and 7 yield nearly identical  
rates for the fcc lattice (Fig.\ \ref{fig:pyc}).  
  
Schramm and Koonin \cite{Schramm} also took into account the  
dynamical effect of motion of the surrounding ions in response to the motion  
of tunneling nuclei in the relaxed lattice (models 4 and 8). This  
effect was described by introducing the effective mass of the  
reacting nuclei. The effective mass appears to be noticeably  
higher than the real nucleus mass, reducing the tunneling  
probability. It turns out that the reduction almost exactly  
compensates the increase of the tunneling probability due to the  
lattice relaxation neglecting the  
effective mass effects \cite{svh69}.  
Accordingly, model 4 of Schramm and Koonin  
\cite{Schramm} gives almost the same reaction rate  
as model 1 (for bcc); and model 8 gives almost the same  
rate as model 5 (for fcc).  
This means that the two limiting approximations,  
the static-lattice and relaxed-lattice, yield very similar  
reaction rates. It is natural to expect that the actual reaction  
rate (to be calculated for the dynamically responding lattice)  
would be the same, and the problem of dynamical tunneling is thus  
solved \cite{Schramm}. Notice that this conclusion is made using  
the curvature corrections ${\rm e}^K$ adopted above (whereas the  
accurate curvature correction for the effective mass model has not been  
calculated).  
  
Zero-temperature pycnonuclear reactions in bcc crystals were also  
studied by Ogata, Iyetomi, and Ichimaru \cite{oii91} and Ichimaru,  
Ogata, and Van Horn \cite{iovh92} using MC lattice screening  
potentials. These authors considered one-component and  
two-component ion systems. Model 9 of Table \ref{tab:pyc}  
represents their results for one-component bcc crystals. In order  
to calculate the tunneling probability, the authors used the  
mean-force plasma screening potential $H(r)$, obtained from MC  
sampling in a classical bcc crystal at $r \gtrsim  
a$ and extrapolated to $r \to 0$ (see Section \ref{thermoscreen}).  
This potential is static and spherically symmetric. It cannot take  
into account the dynamics of lattice response and the anisotropic  
character of the real screening potential in a lattice.  
Furthermore, the barrier penetration was calculated by solving  
numerically the effective radial Schr\"odinger equation. This  
procedure is more approximate than the direct WKB approach of  
Salpeter and Van Horn \cite{svh69} and Schramm and Koonin  
\cite{Schramm} (particularly, it neglects the curvature  
corrections). Numerically, Ichimaru, Ogata and  
Iyetomi \cite{oii91} give a reaction rate which is close to the  
relaxed lattice model (model 2) of Salpeter and Van Horn  
\cite{svh69}. The main reason for the coincidence of these rates  
is that the screening potential in the radial equation of  
Ichimaru, Ogata and Iyetomi is close to the relaxed-lattice  
screening potential of Salpeter and Van Horn at ion separations  
$r\sim 1.5 \ a$, most important for pycnonuclear tunneling problem  
(see Fig.\ 2 in Ref.\ \cite{oii91}).  
  
In spite of the differences in theoretical models 1--9, they result  
in similar reaction rates (Fig.\ \ref{fig:pyc}). According to the above  
discussion, models 1 and 4 seem to be the most reliable among all  
available models for the bcc lattice, while models 5 and 8  
seem to be the most reliable for fcc. These reaction rates may be modified,  
for instance, by taking into account the quantum effect of the spreading  
of WKB trajectories or by a more careful treatment of the  
center-of-mass motion of reacting nuclei. Such effects will possibly  
reduce the reaction rate (as discussed in Ref.\ \cite{pm04} with  
regard to the spreading of WKB trajectories). This could have been  
studied by direct PIMC simulations (e.g., Refs.\  
\cite{ogata97,pm04}). PIMC is also a good tool to confirm the  
conclusions on dynamical effects of lattice response. However,  
PIMC is time consuming and requires very powerful computers. It is  
not clear whether today's computer capabilities are sufficient to  
obtain accurate PIMC pycnonuclear reaction rates.  
  
We suggest to calculate the reaction rates from  
Eq.\ (\ref{pyc-numer}) taking into account that  
the constants $C_{\rm exp}$, $C_{\rm pl}$ and $C_{\rm pyc}$ are  
not known very precisely. In particular, we propose two  
``limiting'' purely phenomenological sets of these constants  
labeled as models 10 and 11 in Table \ref{tab:pyc}. These limiting  
parameters define 
the
maximum and minimum reaction rates which enclose  
all model reaction rates 1--4 and 9 (proposed in the literature  
for the bcc lattice in a density range where the pycnonuclear  
carbon burning is important). They also enclose the most reliable  
models 5 and 8 for the fcc lattice.  
  
The crucial parameter for modeling pycnonuclear fusion is the  
exponent $F_{\rm pyc}=\exp(-C_{\rm pyc}/\sqrt{\lambda})$ in Eq.\  
(\ref{Expyc}) that characterizes the probability of Coulomb  
tunneling. It is easy to show that the exponent argument behaves  
as $C_{\rm pyc}/\sqrt{\lambda} = \alpha\, (r_{12}/r_{\rm qm})^2  
\propto \rho^{-1/6}$, where $r_{12}$ is the equilibrium distance  
between the interacting nuclei in their lattice sites, $r_{\rm  
qm}$ is the rms displacement of the nucleus due to zero-point  
vibrations in its lattice site, and $\alpha \sim 1$ is a numerical  
factor which depends on a model of Coulomb tunneling. The usual  
condition is $r_{\rm qm} \ll r_{12}$ (and the tunneling length  
$\gg r_{\rm qm}$). The exponent argument  
$C_{\rm pyc}/\sqrt{\lambda}$  
is typically large but  
decreases with growing $\rho$, making the Coulomb barrier more  
transparent. The tunneling is actually possible for closest neighbors  
(smallest $r_{12}$); the tunneling of more distant nuclei (higher  
$r_{12}$) is exponentially suppressed. Elastic lattice properties  
specify $r_{\rm qm}$ and $\alpha$, and are, thus, most important  
for the reaction rate. The presence of different ion species,  
lattice impurities and imperfections may drastically affect the  
rate \cite{svh69}.  
  
\subsection{Thermally enhanced pycnonuclear regime}  
\label{thermopyc}  
  
The thermally enhanced pycnonuclear burning occurs with increasing  
$T$; it operates \cite{svh69} in a relatively narrow  
temperature interval $0.5\, T_p / \ln(1/\sqrt{\lambda})  
 \lesssim T \lesssim  0.5 \, T_p$. In this interval  
the majority of the nuclei occupy their ground states in a  
strongly coupled quantum Coulomb system, but the main contribution  
to the reaction rate comes from a 
tiny fraction
of nuclei which  
occupy excited bound energy states. The  
increase of the excitation energy 
increases the penetrability of the Coulomb barrier, 
and makes the excited states  
more efficient 
than the ground state.  
  
The thermally enhanced pycnonuclear regime has been  
studied less accurately than the zero-temperature pycnonuclear regime.  
Salpeter and Van Horn \cite{svh69} calculated the  
thermally enhanced pycnonuclear reaction rate for models 1 and 2  
of a bcc lattice in the WKB approximation. The spectrum of excited  
quantum states was determined for a relative motion of interacting  
nuclei in an anisotropic harmonic oscillator field; the summation  
over discrete quantum states in the expression for the reaction  
rate was replaced by 
the
integration.  
According to their Eq.\ (45), the enhancement of the reaction rate is  
approximately described by  
\begin{equation}  
    { R_{\rm pyc}(T) \over R_{\rm pyc}(0)} -1 =  
     {\Omega \over  \lambda^{1/2} } \,  
     \exp  
     \left(  
         - \Lambda \,{T_p  \over T} +  
         { \Omega_1 \over \sqrt{\lambda}}\,  
        {\rm e}^{ - \Lambda_1 T_p / T}  
     \right),  
\label{pyctherm}  
\end{equation}  
where $\Omega$, $\Omega_1$, $\Lambda$,  
and $\Lambda_1$ are model-dependent  
dimensionless constants.  
The exponent  
$\exp (-\Lambda \,T_p  / T)$  
reflects the Boltzmann  
probability to occupy excited quantum states  
while the double exponent  
$\exp \{ (\Omega_1 / \sqrt{\lambda})\,  
{\rm e}^{ - \Lambda_1 T_p / T} \}$  
describes the enhancement itself.  
In this case the  
characteristic energy of the reacting nuclei is  
\begin{equation}  
  E_{\rm pk} \approx C_1 \hbar  \omega_p +  
  C_2 \, { Z^2 e^2 \over a} \,  
  \exp \left(-\Lambda_1 \,{ T_p \over T }\right),  
\label{pk-pyctherm}  
\end{equation}  
where $C_1$ and $C_2$ are new dimensionless  
constants ($\sim 1$). When $T$ increases from $T=0$ to  
$T \sim 0.5\,T_p$, the characteristic energy  
$E_{\rm pk}$ increases from the ground state level,  
$E_{\rm pk} \sim \hbar \omega_p$, to the top of the Coulomb potential well,  
$E_{\rm pk}  \sim Z^2 e^2/a$.  
  
The thermally enhanced pycnonuclear  
reaction rate was studied also by Kitamura and Ichimaru \cite{ki95} adopting  
the formalism of Ogata, Iyetomi and Ichimaru \cite{oii91}  
(Sections \ref{thermoscreen} and \ref{pycnozero}). The relative  
motion of interacting nuclei was described by a model radial  
Schr\"odinger equation which employed the angle-averaged static MC  
plasma screening potential. The excited energy states were  
determined from the solution of this equation. Such an approach  
seems to be oversimplified. It gives the temperature dependence of  
the reaction rate (Eqs.\ (14) and (15) in Ref.\ \cite{ki95})  
which, functionally, differs from the temperature dependence,  
Eq.\ (\ref{pk-pyctherm}), predicted by Salpeter and Van Horn.  
Nevertheless, numerically, both temperature dependencies at $T  
\lesssim 0.5\, T_p$ are in a reasonable qualitative agreement.  
  
We expect that the reaction rate in the thermally enhanced  
pycnonuclear regime will be further  
elaborated in the future.  
  
\subsection{The intermediate thermo-pycnonuclear regime}  
\label{intermed}  
  
The intermediate thermo-pycnonuclear regime is realized at  
temperatures $T \sim T_p$ (roughly, at $T_p/2 \lesssim T \lesssim  
T_p$) which separate the domains of quantum and classical ion  
systems. The main contribution to the reaction rate stems then from  
nuclei which are either slightly bound, or slightly unbound, with respect  
to their potential wells. The calculation of the reaction rate in  
this regime is complicated. We will describe this rate by a  
phenomenological expression presented in the following section.  
  
\subsection{Single analytical approximation in all regimes}  
\label{together}  
  
Let us propose a phenomenological expression for the reaction rate which  
combines all the five burning regimes:  
\begin{eqnarray}  
    R &= & R_{\rm pyc}(T=0) + \Delta R(T), \quad  
    \Delta R(T) =   { n_i^2 \over 2} \, S(E_{\rm pk}) \,  
     { \hbar \over mZ^2e^2 }\, P \, F,  
\nonumber \\  
     F &= &\exp \left(-\widetilde{\tau}  
     +C_{\rm sc}\widetilde{\Gamma}\,  
     \varphi \,{\rm e}^{-\Lambda T_p/T}  
     -\Lambda\, {T_p \over T} \right),  
     \quad  
     P= { 8 \, \pi^{1/3} \over \sqrt{3}\; 2^{1/3}} \,  
       \left(E_{\rm a} \over \widetilde{T} \right)^\gamma.  
\label{overall}  
\end{eqnarray}  
In this case, $\varphi = \sqrt{\Gamma}/[(C_{\rm  
sc}^4/9)+\Gamma^2]^{1/4}$;  
$R_{\rm pyc}(T=0)$ is the zero-temperature pycnonuclear reaction  
rate (Section \ref{pycnozero}); $\Delta R(T)$  
is the temperature-dependent part (with a product of an  
exponential function $F$ and a pre-exponent $P$). The quantities  
$\widetilde{\tau}$ and  $\widetilde{\Gamma}$ are  
similar to the familiar quantities $\tau$ and $\Gamma$, but  
contain a ``renormalized'' temperature $\widetilde{T}$:  
\begin{equation}  
    \widetilde{\tau}=3\, \left(\pi \over 2 \right)^{2/3}  
    \left(E_{\rm a} \over  
     \widetilde{T} \right)^{1/3},  
    \quad  
    \widetilde{\Gamma} = {Z^2 e^2 \over a \widetilde{T} },  
    \quad  
    \widetilde{T} = \sqrt{ T^2+C_T^2T_p^2},  
\label{renorm}  
\end{equation}  
where $C_T$ is a dimensionless  
renormalization parameter ($\sim 1$). For high temperatures $T  
\gg T_p$ we have $\widetilde{\tau} \to \tau$, $\widetilde{\Gamma}  
\to \Gamma$, and $\widetilde{T} \to T$. In this case the  
temperature dependent term tends to $\Delta R(T) \to R_{\rm th}(T)  
\gg R_{\rm pyc}$, and Eq.\ (\ref{overall}) reproduces the  
thermonuclear reaction rate (Sections\ \ref{thermo} and  
\ref{thermoscreen}). At low temperatures $T \la T_p$ the  
quantities $\widetilde{\tau}$, $\widetilde{\Gamma}$ and  
$\widetilde{T}$, roughly speaking, contain ``the quantum  
temperature'' $T_p$ rather than the real temperature $T$ in the  
original quantities $\tau$, $\Gamma$ and $T$.  
In the limit of $T \to 0$ we obtain  
$\widetilde{\Gamma}=1/[\sqrt{\lambda}\,(72 \pi)^{1/6} \, C_T]$  
and $\widetilde{\tau}=\left(3 \, \sqrt{\pi/\lambda} \, \right)/  
\left(2^{7/6} \,C_T^{1/3} \, \right)$.  
  
At this point, let us require that at $T \ll T_p$ the factor  
$\exp(-\widetilde{\tau})$ in the exponential  
function $F$, Eq.\ (\ref{overall}), reduces to  
$\exp(-C_{\rm exp}/\sqrt{\lambda})$ in the exponential function 
$F_{\rm pyc}$,  
Eq.\ (\ref{Expyc}). This would allow us to obey Eq.\ (\ref{pyctherm}) by  
satisfying the equality  
\begin{equation}  
   {3\, \sqrt{\pi} /( 2^{7/6}\, C_T^{1/3})}  
   =C_{\rm exp}.  
\label{Crelat}  
\end{equation}  
Taking $C_{\rm exp}$ we can determine $C_T$ (see Table \ref{tab:pyc}). The  
double exponent factor in $F$, Eq.\ (\ref{overall}), will correspond to the  
double exponent factor in  Eq.\ (\ref{pyctherm}).  
Strictly speaking, Eq.\ (\ref{pyctherm})  
contains two different constants $\Lambda$ and  
$\Lambda_1$. However, taking into account the  
uncertainties of $R$ in the thermally enhanced pycnonuclear regime  
(Section \ref{thermopyc}), we replace two constants by one.  
  
Finally, the quantity $\gamma$ in Eq.\ (\ref{overall})  
should be taken in such a way as to reproduce the correct limit  
$\gamma_1=2/3$ at $T \gg T_p$ (Sect.\ \ref{thermo})  
and $\gamma_2=(2/3)\,(C_{\rm pl}+0.5)$ at  
$T \ll T_p$ (see Eq.\ (\ref{pyctherm})).  
The natural interpolation expression for $\gamma$ would be  
\begin{equation}  
    \gamma=(T^2 \gamma_1 + T_p^2 \gamma_2)/(T^2+T_p^2).  
\label{gamma}  
\end{equation}  

In addition, we need the reaction energy $E_{\rm pk}$ to evaluate  
the astrophysical factor $S(E_{\rm pk})$.  
Since the $S$-factor is a slowly varying function of energy,  
it is reasonable to approximate  $E_{\rm pk}$ by the expression  
\begin{equation}  
    E_{\rm pk}= \hbar \omega_p  
    + \left( {Z^2 e^2 \over a} + {T \tau \over 3} \right) \,  
    \exp \left( - { \Lambda \, T_p \over T} \right)  
\label{peak}  
\end{equation}  
which combines the expressions in the thermonuclear and  
pycnonuclear regimes. To avoid the introduction of  
many fit parameters (unnecessary at the  
present state of investigation),  
we set $C_1=C_2=1$ in Eq.\ (\ref{pk-pyctherm}).  
  
Thus, we propose to adopt the analytic expression (\ref{overall})  
using the following parameters:\\  
    (1) $C_{\rm sc}=1.0754$ for the  
case of strong plasma screening in the thermonuclear regime  
(Section~\ref{thermoscreen});\\  
    (2) $C_{\rm exp}$, $C_{\rm pyc}$,  
and $C_{\rm pl}$ for conditions of zero-temperature pycnonuclear  
burning
(see Table~\ref{tab:pyc} and Section~\ref{pycnozero});\\
    (3) The quantum-temperature constant $C_T$
(Section~\ref{pycnozero}),    
which is important at  
$T \sim T_p$ and expressed through $C_{\rm exp}$ via Eq.\ (\ref{Crelat});  
the corresponding values of $C_T$ are listed in Table~\ref{tab:pyc};\\  
    (4) The 
last 
constant $\Lambda$, that is important at $T \sim T_p$,  
is still free.  
  
\begin{figure}[tbh]  
\begin{center}  
\includegraphics[width=12.0cm]{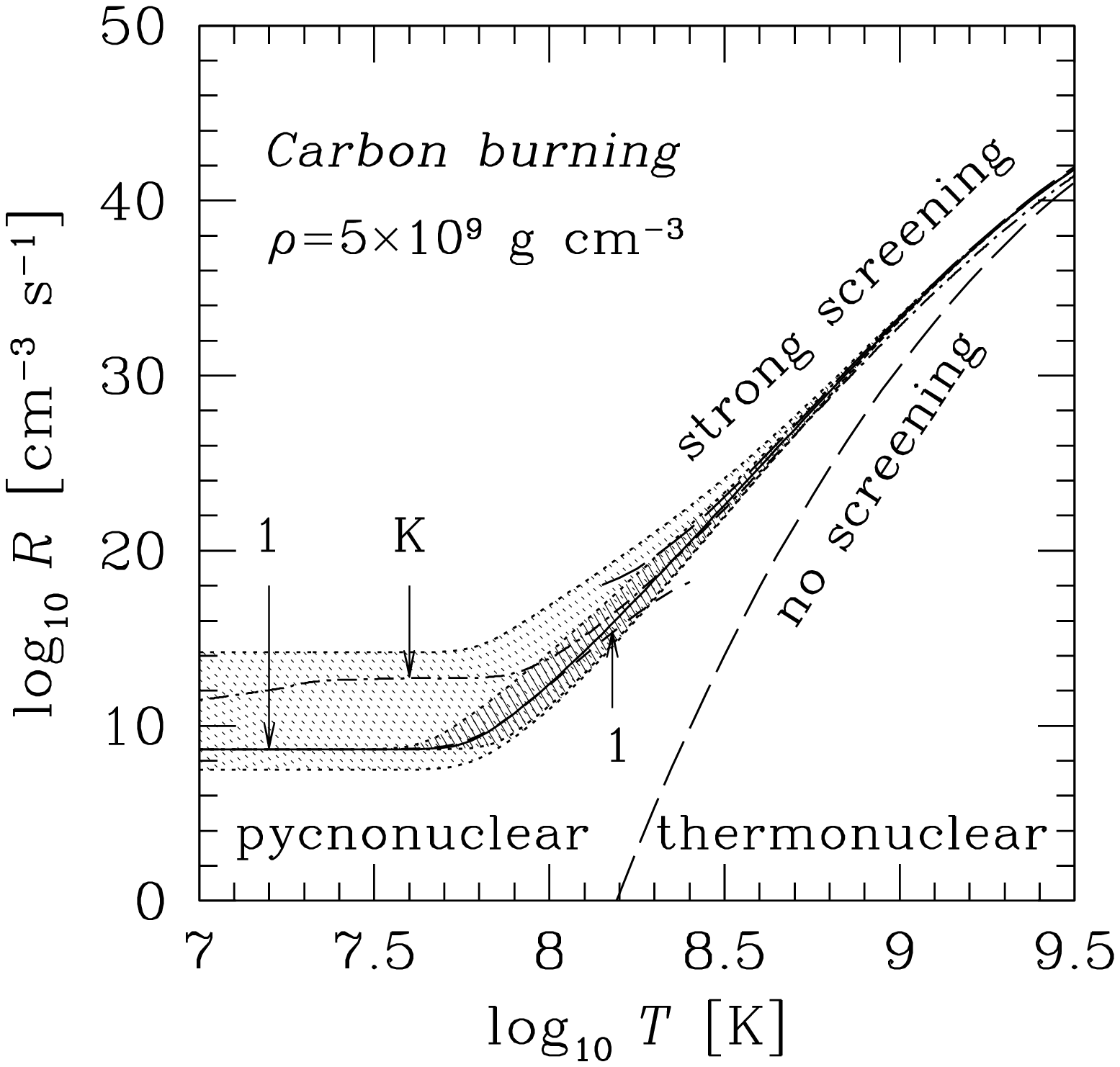}  
\caption{Temperature dependence of the carbon fusion rate at  
$\rho=5 \times 10^9$ g~cm$^{-3}$. The solid line 1 is  
our most optional interpolation expression (Sect.\ \ref{together}),  
based on model 1  from Table \ref{tab:pyc}  
with $\Lambda=0.5$.  
Doubly hatched region shows theoretical  
uncertainties of model 1 associated with variations  
of $\Lambda$ from 0.35 to 0.65.  
The dash-and-dot line `K' is  
the interpolation of Kitamura \cite{kitamura00}.  
The short-dashed line  
1 is calculated from the expressions of Salpeter and  
Van Horn \cite{svh69}, which are valid in the pycnonuclear regime  
($T=0$ and with the thermal enhancement). Long-dashed lines  
show the thermonuclear reaction rates calculated with  
account for plasma screening (Sect.\ \ref{thermoscreen})  
and without screening (Sect.\ \ref{thermo}).  
Singly hatched region displays total assumed theoretical uncertainties  
of the reaction rates.  
}  
\label{fig:therm}  
\end{center}  
\end{figure}

We have checked (Fig.\ \ref{fig:therm}) that taking  
the optional model 1 from Table \ref{tab:pyc}  
and the value $\Lambda=0.5$  
results in a good agreement  
with the carbon burning rate  
calculated at $\rho \sim 10^9-10^{10}$  
g~cm$^{-3}$ and $T \lesssim 0.5 \, T_p$  
from Eq.\ (45) of Salpeter and  
Van Horn \cite{svh69}  (for the  
thermally enhanced pycnonuclear regime).  
(Notice that model 2 requires slightly  
lower $\Lambda \approx 0.45$.)  
Taking $\Lambda=0.35$ leads to a noticeably  
higher reaction rate at $T \lesssim 0.5\, T_p$,  
while taking $\Lambda=0.65$ leads to a noticeably  
lower rate.  
  
Accordingly, for any model  
of zero-temperature pycnonuclear burning  
from Table \ref{tab:pyc} we suggest to adopt $\Lambda=0.5$ as optional,  
$\Lambda=0.35$ to maximize and $\Lambda=0.65$ to minimize  
the reaction rate. In particular, model 1 with $\Lambda=0.5$  
seems to be the ``most optional'';  
our limiting model 10 from  
Table \ref{tab:pyc} with $\Lambda=0.35$ is expected to  
give the upper theoretical limit for the reaction rate,  
while the other limiting  
model 11 with $\Lambda=0.65$ is expected to give the  
lower theoretical limit. We also need  
the astrophysical factor $S(E)$,  
which was described in Section \ref{sfactor} for the carbon burning.  
We could  easily  
introduce additional constants to tune our phenomenological  
model when precise  
calculations of reaction rates appear in the future.  
  
For illustration, Fig.\ \ref{fig:therm} shows  
the temperature dependence of the carbon burning rate  
at $\rho=5 \times 10^9$~g~cm$^{-3}$.  
The solid curve is the  
most optimal model (based on both -- zero-temperature  
and thermally enhanced -- pycnonuclear burning models  
of Salpeter and Van Horn \cite{svh69} for the bcc static lattice).  
The double hatched region shows  
assumed uncertainties of this model associated with  
variations of $\Lambda$ from 0.35 to 0.65 (as if we accept  
the zero-temperature model but question the less elaborated model of thermal  
enhancement).  
The singly hatched region indicates overall uncertainties  
(limited by the models of the maximum and minimum reaction rates).  
The lower long-dashed line is obtained assuming classical  
thermonuclear burning without any screening (Section \ref{thermo}).  
The upper long-dashed line  
is calculated using the formalism of thermonuclear burning  
with screening (Section \ref{thermoscreen}). The  
screening enhancement of the  
reaction rate becomes stronger with the decrease of $T$.  
The formalism for describing this enhancement  
is expected to be valid at $T \gtrsim T_p$,  
but we intentionally extend the upper long-dashed  
curve to $T=0.5\,T_p$, where  
the formalism breaks down and the curve diverges from  
the expected (solid) curve. The short-dash curve is  
calculated from the equations of Salpeter and Van Horn  
\cite{svh69} derived in the thermally enhanced pycnonuclear  
regime (model 1) and valid at $T \lesssim 0.5\, T_p$.  
We intentionally extend the curve to higher $T$,  
where the formalism of thermally enhanced pycnonuclear burning  
becomes invalid and the curve diverges  
from the expected curve. Our phenomenological solid curve  
provides a natural interpolation  
at $T \sim T_p$ between the short-dashed curve  
and the upper long-dashed curve.  
  
More complicated expressions for the reaction rate $R$  
in wide ranges of $\rho$ and $T$ were proposed  
by Kitamura \cite{kitamura00}  
who combined the results of recent calculations of $R$  
in the different regimes. His expressions are mainly  
based on the results of Refs.\ \cite{oii91,iovh92,oiv93,ki95,ogata97}  
which are not free of approximations  
(as discussed in Sections \ref{thermoscreen}, \ref{pycnozero},  
and \ref{thermopyc}).  
In contrast to our formula,  
Kitamura took into account the effects of electron screening  
(finite polarizability of the electron gas) and considered  
the case of equal and non-equal reacting nuclei.  
However, the electron screening effects are relatively  
weak; their strict inclusion in the pycnonuclear regime  
represents a complicated problem.  
We do not include them but, instead, take into account  
theoretical uncertainties of the reaction rates without  
electron screening. We have checked that the results  
by Kitamura \cite{kitamura00} for carbon burning in the most important  
$T-\rho$ domain lie well within these uncertainties.  
  
Our formula gives a smooth  
behavior of the reaction rate as a function of  
temperature and density, without 
any jump 
at the melting  
temperature $T=T_{\rm m}$.  
We do not expect any strong 
jump 
of such kind since  
the liquid-solid phase transition in dense stellar matter is tiny.  
A careful analysis shows the absence  
of noticeable jumps of transport coefficients  
\cite{baikoetal98}
and the neutrino emissivity  
owing to electron-nucleus bremsstrahlung. 
A direct example is given by  
the theory of nuclear burning. Ichimaru and Kitamura  
\cite{ik99} predicted a noticeable jump of the  
reaction rate at $T=T_{\rm m}$, while a more careful  
analysis of Kitamura \cite{kitamura00} considerably  
reduced this jump.  
  
\section{Carbon burning and ignition in dense matter}  
\label{carbonburn}  
  
In this section we will analyze the rate of the $^{12}$C+$^{12}$C  
reaction as a function of $T$ and $\rho$ and investigate the  
conditions for carbon burning in dense stellar matter.  
  
Because
the probability for Coulomb tunneling depends exponentially  
on plasma parameters, changes in density $\rho$ and temperature  
$T$ have dramatic effects on the burning rate $R$. In  
thermonuclear regimes (Sections \ref{thermo} and  
\ref{thermoscreen}) the $^{12}$C+$^{12}$C rate is more sensitive to  
changes in temperature $T$ than in density $\rho$. On the  
contrary, in pycnonuclear regimes (Sections \ref{pycnozero} and  
\ref{thermopyc}) the rate depends significantly on the density  
$\rho$.  For instance, if $T$ decreases from $3 \times  
10^9$~K to $3 \times 10^8$~K at $\rho= 5 \times 10^9$  
g~cm$^{-3}$ (Fig.\ \ref{fig:therm}; thermonuclear burning with  
strong screening), the reaction rate drops by $\sim 20$ orders of  
magnitude.  
Neglecting the enhancement due to plasma screening, the  
rate will drop by ten more orders of magnitude. An increase in  
density $\rho$ from $10^8$ g~cm$^{-3}$ to $10^{11}$ g~cm$^{-3}$ at  
$T \la 3 \times 10^7$ K (in the zero-temperature pycnonuclear  
regime) results in a rate increase of $\sim 100$ orders of  
magnitude (Fig.\ \ref{fig:pyc}). Note that no carbon can survive  
in a degenerate matter at $\rho>3.90 \times 10^{10}$ g~cm$^{-3}$  
because of the double electron capture ${}^{12}_{~6}{\rm C} \to  
{}^{12}_{~5}{\rm B} \to {}^{12}_{~4}{\rm Be}$ (e.g., Shapiro and  
Teukolsky \cite{st83}). The electron capture has a well defined  
density threshold, $3.9 \times 10^{10}$ g~cm$^{-3}$, and proceeds  
quickly after the threshold is exceeded. We will ignore this  
process in the present section.  
  
The strong dependence of the rate $R$ on density $\rho$ and  
temperature $T$ leads to huge variations of the characteristic  
time scale $\tau_{\rm burn}=n_i/R$ for carbon burning. Figure  
\ref{fig:diag} shows two solid lines in the $\rho-T$ plane, along which  
$\tau_{\rm burn}=1$~s and $10^{10}$ years  
(nearly the Universe age), respectively.  
They are calculated using the most optional carbon burning  
model (model 1 from Table \ref{tab:pyc}, $\Lambda=0.5$).  
The lines are almost horizontal in the  
thermonuclear burning regime ($R$ is a  
slowly varying function of $\rho$) and almost vertical in the  
pycnonuclear regime ($R$ is a slowly varying function of $T$).  
The bending part of the  lines  
corresponds to the thermally enhanced  
pycnonuclear and intermediate thermo-pycno nuclear regimes.  
At  $T$ and $\rho$ above the upper line the  
burning time is even shorter  
than 1 s; at these conditions no carbon will  
survive in the dense matter of astrophysical objects.  
For conditions below the lower solid line, $\tau_{\rm burn}$ is longer  
than $10^{10}$ years, and carbon burning can be  
disregarded for most applications.  
  
Thus, the 
studies of 
carbon  
burning 
can be focused on
the narrow strip in the $\rho-T$ plane  
between the lines of $\tau_{\rm burn}=1$ s and $\tau_{\rm  
burn}=10^{10}$ yr. 
Hatched regions show theoretical uncertainties  
of each line (limited by the maximum and minimum reaction  
rate models, Section \ref{together}).  
The uncertainties are relatively small  
in the thermonuclear regime where all models give  
nearly the same reaction rate. The uncertainties are higher in other  
burning regimes.  
  
Having a carbon burning model we can plot  
the carbon ignition curve. This curve is a necessary ingredient  
for modeling nuclear explosions of massive white dwarfs  
(producing supernova Ia events, so important for cosmology;  
see, e.g., Refs.\ \cite{SNI,filippenko04})  
and for modeling carbon explosions of matter in accreting  
neutron stars (viable models of superbursts observed recently  
from some accreting neutron stars; e.g.,  
Refs.\  \cite{CB01,SB02,cn04}).  
  
The ignition curve is commonly determined as the line in the  
$\rho-T$ plane (Fig.\ \ref{fig:diag}) where  
the nuclear energy generation rate is equal to the local  
neutrino emissivity of dense matter (the neutrino emission carries  
the generated energy out of the star). At higher $\rho$ and $T$  
(above the curve) the nuclear energy generation rate exceeds the  
neutrino losses and carbon ignites. In Fig.\ \ref{fig:diag} we  
present the carbon ignition (solid) curve,  
calculated using the most optional model of carbon burning,  
together with its uncertainties (limited by the minimum and  
maximum rate models). The neutrino energy losses are assumed  
to be produced by plasmon decay and by electron-nucleus  
bremsstrahlung. The neutrino emissivity due to plasmon decay is  
obtained from extended tables calculated by M.~E.~Gusakov  
(unpublished); they are in good agreement with the results by Itoh  
et al.\ \cite{itohetal92}. The neutrino bremsstrahlung emissivity  
is calculated using the formalism of Kaminker et al.\  
\cite{kaminkeretal99}, which takes into account electron band  
structure effects in crystalline matter.  
  
For $\rho \la  10^9$ g~cm$^{-3}$ theoretical uncertainties  
of the  ignition curve are seen to be small.  
They become  
important at $\rho \ga 10^9$ g~cm$^{-3}$ and $T \sim (1-3)\times  
10^8$ K in the intermediate thermo-pycnonuclear burning regime and  
the thermally enhanced burning regime. This $\rho-T$ range is  
appropriate for central regions of massive and warm white dwarfs  
which may produce type Ia supernova explosions. Lower $T$ are also  
interesting for these studies (e.g., Baraffe et al.\  
\cite{bhw04}).  
  
If we formally continue the ignition curve to lower $T$, it  
will bend and shift to lower densities, where the nuclear burning  
time scale $\tau_{\rm burn}$ is exceptionally slow exceeding the  
age of the Universe. The bend is associated with a very weak  
neutrino emission at $T \la 10^8$ K. These parts of the ignition  
curve are oversimplified because at low  
$T$ the energy outflow produced by thermal conduction  
becomes more efficient than the outflow due to the neutrino  
emission. These parts are shown by the long-dash line  
(and their uncertainties are indicated by thin dash-and-dot lines).  
Unfortunately, the conduction energy outflow is  
non-local and ``non-universal''. It depends on specific conditions  
of the burning environment (a white dwarf core or a neutron star  
crust) and the associated thermal conductivity (provided mainly by  
strongly degenerate electrons). In this case the ignition becomes  
especially complicated. A very crude estimate shows that the  
ignition curve, governed by the thermal conduction, is nearly  
vertical and close to the $\tau_{\rm burn}=10^{10}$ yr curve in  
the range of $T$ from $10^8$~K to $10^6$~K  
in Fig.\ \ref{fig:diag}. At  $T\la 10^6$~K the  
curve is strongly affected by the thermal conductivity model. In a  
cold ideal carbon crystal, umklapp processes of electron-phonon  
scattering are frozen out (e.g., Ref.\ \cite{ry82}).  
Under these conditions the electron conduction is determined by  
inefficient normal electron-phonon scattering, leading to high  
conductivity values. This shifts the ignition to higher $\rho$. On  
the other hand, carbon matter may contain randomly located ions of  
other elements (charged impurities) which can keep the electron  
Coulomb scattering rather efficient and maintain a low electron  
thermal conductivity. In this case the ignition curve at $T \la  
10^6$~K remains nearly vertical.  
  
\section{Conclusion}  
\label{conclusions}  
  
The goal of this paper was to develop a phenomenological formalism  
for calculating fusion reaction rates between identical nuclei.
This formalism should be applicable for a broad range of thermonuclear and  
pycnonuclear burning scenarios. It involves a generalized treatment for 
calculation of the fusion probability at low energies and the development of 
a single simple phenomenological expression for the fusion rate  
valid in a wide range of temperatures and densities.
  
We have introduced a generalized model approach for calculating the 
$S$-factor of heavy-ion fusion reactions relevant
for stellar nucleosynthesis processes. We  
have demonstrated the applicability and reliability of the  
approach by calculating the astrophysical factor $S(E)$ for the  
carbon fusion reaction $^{12}$C+$^{12}$C (Section \ref{sfactor})  
and by comparing the theoretical results with experimental data.  
  
Furthermore, we
have analyzed (Section \ref{fusion}) previous  
calculations of the fusion rate for identical nuclei in stellar  
matter, with emphasis on the complicated problem of Coulomb  
barrier penetration in a dense-plasma environment. Combining the  
results of previous studies, we have proposed a single simple  
phenomenological expression for the fusion rate, valid in all five  
fusion regimes (that can be realized in the different $\rho-T$  
regions). Our formula contains adjustable parameters whose  
variations reflect theoretical uncertainties of the reaction  
rates.  
  
For illustration, we have considered (Section \ref{carbonburn}) the efficiency 
of carbon burning in dense matter and the conditions for carbon ignition  
in white dwarf cores and neutron star crusts. We show that carbon  
burning is actually important in a sufficiently narrow $\rho-T$  
strip which is mainly determined by the temperature 
$T \sim (4-15) \times 10^8$ K 
as long as 
$\rho \lesssim  3 \times 10^9$  
g~cm$^{-3}$, and by the density 
$\rho \sim (3-50) \times 10^9$  g~cm$^{-3}$ 
as long as $T \lesssim 10^8$~K. On the basis of these  
results we suggest that the current knowledge of nuclear fusion is  
sufficient to understand the main features of carbon burning in  
stellar matter,
especially at $\rho \lesssim 3 \times 10^9$~g~cm$^{-3}$.  
  
We have focused on the simplest case of heavy-ion burning in a one-component  
Coulomb system; particularly, in a perfect crystal. There is no  
doubt that dense matter of white dwarfs and neutron stars are more  
complicated and require a more complex approach taking into  
account mixtures of different heavy nuclei and imperfections in  
dense matter. The complexity ranges from essentially two-component plasma 
conditions anticipated in the carbon-oxygen cores
of white dwarfs to the multi-component  
isotope distribution in the ashes of accreting neutron stars  
\cite{woos04}.  
  
In a forthcoming paper we will expand the presented 
analysis to the case of the fusion rates between different isotopes. We will 
employ this formalism for calculating the $S$-factors for a broad range 
of heavy-ion fusion reactions. We will include the results in
a pycno-thermonuclear reaction network and simulate the nucleosynthesis 
in high density stellar matter.

\begin{acknowledgments}  
We are grateful to H.~DeWitt for critical remarks and to  
M.~Gusakov for providing the tables of neutrino emissivities due to  
plasmon decay. This work was partially supported by The Joint Institute for  
Nuclear Astrophysics (JINA) NSF PHY 0216783, Funda\c{c}\~{a}o de Amparo \`a  
Pesquisa do Estado de S\~ao Paulo (FAPESP), CONACYT (M\'exico),  
DoE grant DE-F05-96ER-40983 and BMBF (Germany), under the project 06 MT 193,  
RFBR (grants 03-07-90200 and 
05-02-16245) 
and RLSSP (project 1115.2003.2).  
\end{acknowledgments}

  
  

\newpage  
%
%
\begin{table}[t]  
\caption[]{The table presents the coefficients $C_{\rm exp}$, $C_{\rm pyc}$  
and $C_{\rm pl}$ of the pycnonuclear reaction rate obtained at $T=0$  
(see Eq. \ref{pyc-numer}).  
The dimensionless parameter $C_T$, which is related to the ``renormalized''  
temperature (Eq. \ref{renorm}), is also included in the table.}  
\label{tab:pyc}  
\begin{center}  
\begin{tabular}{|c|c|c|c|c|l|l|}  
\hline  
\hline  
~No.~& ~~~$C_{\rm exp}$~~~  & ~~$C_{\rm pyc}$~~  &~~ $C_{\rm pl}$~~~ &  
~~$C_T$~~~ & ~Model   & Refs. \\  
\hline \hline  
1& 2.638 & 3.90  & 1.25  & 0.724  & bcc; static lattice &  
\cite{svh69, Schramm} \\  
\hline  
2& 2.516 & 4.76  & 1.25  & 0.834  & bcc; relaxed lattice -- WS &  
\cite{svh69,Schramm} \\  
\hline  
3&2.517 & 4.58$^{a)}$  &  1.25  & 0.834 & bcc; relaxed lattice &  
\cite{Schramm} \\  
\hline  
4&2.659 & 5.13$^{a)}$  &  1.25  & 0.707 & bcc; effective mass approx. &  
\cite{Schramm} \\  
\hline  
5&2.401 & 7.43$^{a)}$  &  1.25  & 0.960 & fcc; static lattice &  
\cite{Schramm} \\  
\hline  
6&2.265 & 13.5$^{a)}$ &  1.25  & 1.144 & fcc; relaxed lattice -- WS &  
\cite{Schramm} \\  
\hline  
7&2.260 & 12.6$^{a)}$   &  1.25  & 1.151 & fcc; relaxed lattice &  
\cite{Schramm} \\  
\hline  
8&2.407 & 13.7$^{a)}$  &  1.25  & 0.953 & fcc; effective mass approx. &  
\cite{Schramm} \\  
\hline  
9&2.460 & 0.00181  &  1.809  & 0.893 & bcc; MC calculations &  
\cite{oii91} \\  
\hline  
\hline  
10&2.450 & 50  &  1.25  & 0.904 & maximum rate & present paper \\  
\hline  
11&2.650 & 0.5  &  1.25  & 0.711 & minimum rate & present paper \\  
\hline  
\hline  
\end{tabular}  
\end{center}  
$^{a)}$ Corrected for the curvature factor as explained in the text.  
\end{table}  
  
\begin{thebibliography}{222}  
%
\bibitem{Wallerstein}  
G. Wallerstein et al., Rev. Mod. Phys. {\bf 69}, 995 (1997).  
  
\bibitem{Barnes}  
C. A. Barnes, S. Trentalange and S.-C. Wu, in  
{\em Treatise on Heavy Ion Science} {\bf 6}, Plenum, New York, 1985.  
  
\bibitem{WW93} T. A. Weaver and S. E. Woosley, Phys. Rep. {\bf227}, 65  
               (1993).  
  
\bibitem{WHR03} S. E. Woosley, Alexander Heger, T. Rauscher and R. D.  
                Hoffman, Nucl. Phys. {\bf A 718}, 3 (2003).  
  
  
\bibitem{SNI-1} W. Hillebrandt and J. C. Niemeyer, Ann. Rev. Astr. Astrophys.  
                {\bf 38}, 191 (2000).  
  
\bibitem{SNI} S.E. Woosley, S. Wunsch and M. Kuhlen, Astrophys. J. {\bf 607},  
921 (2004).  
  
\bibitem{bhw04}  
I.\ Baraffe, A.\ Heger and S.\ E.\ Woosley, Astrophys. J.  
{\bf 615}, L378 (2004).  
  
\bibitem{Salp}  
E. E. Salpeter, Australian J. Phys. {\bf 7}, 373 (1954).  
  
\bibitem{CLL02} R. Cussons, K. Langanke and T. Liolios, Eur.  
                Phys. J. {\bf A15}, 291 (2002).  
  
\bibitem{CB01} A. Cumming and L. Bildsten, Astrophys. J. Lett. {\bf 559}, L127  
               (2001).  
  
\bibitem{SB02} T. E. Strohmayer and E. F. Brown, Astrophys. J.  
               {\bf 566}, 1045 (2002).  
  
\bibitem{cn04}  
R.\ L.\ Cooper and R.\ Narayan, Astrophys. J., 2004,  
submitted [astro-ph/0410462].  
  
\bibitem{SBC03} H. Schatz, L. Bildsten and A. Cumming, Astrophys. J. Lett.  
                {\bf 583}, L87 (2003).  
  
\bibitem{hz}  
P.\ Haensel and L.\ Zdunik, Astron.\ Astrophys.\  
{\bf 229}, 117 (1990); P.\ Haensel and L.\ Zdunik, Astron.\ Astrophys.\  
{\bf 404}, L33 (2003).  
  
\bibitem{svh69}  
E.~E.~Salpeter and H.~M.~Van Horn, Astrophys.\ J.\ {\bf 155}, 183 (1969).  
  
\bibitem{Schramm}  
S.~Schramm and S.~E.~Koonin, Astrophys.\ J.\ {\bf 365}, 296  
(1990); erratum: {\bf 377}, 343 (1991).  
  
\bibitem{ref6} M. A. C\^andido Ribeiro, L. C. Chamon, D. Pereira, M. S.  
               Hussein and D. Galetti, Phys. Rev. Lett. {\bf 78}, 3270 (1997).  
  
\bibitem{ref7} L. C. Chamon, D. Pereira, M. S. Hussein, M. A. C\^andido  
               Ribeiro and D. Galetti, Phys. Rev. Lett. {\bf 79}, 5218 (1997).  
  
\bibitem{ref8} L. C. Chamon, D. Pereira and M. S. Hussein, Phys. Rev.  
               {\bf C58}, 576 (1998).  
  
\bibitem{Toward}  L. C. Chamon, B. V. Carlson, L. R. Gasques, D. Pereira, C.  
                  De Conti,M. A. G. Alvarez, M. S. Hussein, M. A. Candido  
                  Ribeiro, E. S. Rossi Jr. and C. P. Silva, Phys. Rev.  
                  {\bf C66}, 014610 (2002).  
  
\bibitem{Fusion} L. R. Gasques, L. C. Chamon, D. Pereira, M. A. G. Alvarez,  
                 E. S. Rossi Jr., C. P. Silva and B. V. Carlson, Phys. Rev.  
                 {\bf C69}, 034603 (2004), and references therein.  
  
\bibitem{Afanasjev} A. V. Afanasjev, P. Ring and J. K\"{o}nig, Nucl. Phys.  
                   {\bf A676}, 196 (2000).  
  
\bibitem{Afanasjev1} A. V. Afanasjev, J. K\"{o}nig and P. Ring, Phys. Rev.  
                     {\bf C60}, 051303(R) (1999).  
  
\bibitem{Gonzalez} T. Gonzalez-Llarena, J. L. Edigo, G. A. Lalazissis and P.  
                   Ring, Phys. Lett. {\bf B379}, 13 (1996).  
  
\bibitem{NL2} P.-G. Reinhard, M. Rufa, J. Maruhn, W. Greiner and  
              J. Friedrich, Z. Phys. {\bf A323}, 13 (1986).  
  
\bibitem{NL3} G. A. Lalazissis, J. K\"onig and P. Ring,  
              Phys. Rev. {\bf C55}, 540 (1997).  
  
\bibitem{DD-ME1} T. Niksic, D. Vretenar, P. Finelli and P. Ring,  
                 Phys. Rev. {\bf C66}, 024306 (2002).  
  
\bibitem{DD-ME2}  G. Lalazissis et al., in progress.  
  
\bibitem{c12-dens} L. R. Gasques, L. C. Chamon, C. P. Silva, D. Pereira,  
                   M. A. G. Alvarez, E. S. Rossi Jr., V. P. Likhachev, B. V.  
                   Carlson and C. De Conti, Phys. Rev. {\bf C65}, 044314  
                   (2002).  
  
\bibitem{Hill} D. L. Hill and J. A. Wheeler, Phys. Rev. {\bf 89}, 1102  
               (1953).  
  
\bibitem{Schiff} L. I. Schiff, {\em Quantum Mechanics}, 3rd ed. (McGraw-Hill,  
                 New York, 1968).  
  
\bibitem{FCZII} W. A. Fowler, G. R. Gaughlan and B. A. Zimmerman, Annu. Rev.  
                Astro. Astrophys. {\bf 13}, 69 (1975).  
  
  
\bibitem{Patterson} J. R. Patterson, H. Winkler and C. S. Zaidins,  
                    Astrophys. J. {\bf 157}, 367 (1969).  
  
\bibitem{Mazarakis} M. G. Mazarakis and W. E. Stephens, Phys. Rev.  
                    {\bf C7}, 1280 (1973).  
  
\bibitem{High} M. D. High and B. Cujec, Nucl. Phys. {\bf A282}, 181 (1977).  
  
\bibitem{Eli} P. Rosales et al., Rev. Mex. F\'{\i}s. {\bf 49}, 88 (2003).  
  
\bibitem{Kettner} K. U. Kettner, H. Lorenz-Wirzba and C. Rolfs, Z. Phys.  
                  {\bf A298}, 65 (1980).  
  
\bibitem{Becker} H.\ W.\ Becker et al., Z.\ Phys.\ {\bf A303}, 305 (1981).  
  
\bibitem{Oh87} S. Ohkubo and D. M. Brink, Phys. Rev. {\bf C36}, 966 (1987).  
  
\bibitem{Mi72} G. J. Michaud and E. W. Vogt, Phys. Rev. {\bf C5}, 350 (1972).  
  
\bibitem{Treu80} W. Treu, H. Fr\"{o}hlich, W. Galster, P. D\"{u}ck and 
H. Voit, Phys. Rev. {\bf C22}, 2462 (1980).

\bibitem{Al03} M. A. G. Alvarez et al., Nucl. Phys. {\bf A723}, 93 (2003).

\bibitem{dewittetal03}  
H.\ E.\ DeWitt, W.\ Slattery,  
D.\ Baiko and D.\ Yakovlev, Contrib.\ Plasma Phys.\ {\bf 41}, 251 (2001).  
  
\bibitem{aj78}  
A.\ Alastuey and B.\ Jancovici, Astrophys. J. {\bf 226}, 1034  (1978).  
  
\bibitem{ogata97}  
S.\ Ogata, Astrophys.\ J.\ {\bf 481}, 883 (1997).  
  
\bibitem{ys89}  
D.G. Yakovlev and D.A. Shalybkov, Soviet Sci.\ Rev., Sect.\  
{\bf E7}, 313 (1989).  
  
\bibitem{dgc73}  
H.\ E.\ DeWitt, H.\ C.\ Graboske and M.\ S.\ Cooper,  
Astrophys.\ J.\  {\bf 181}, 439 (1973).  
  
\bibitem{rosenfeld96}  
Y.~Rosenfeld, Phys.\ Rev.\ {\bf E53}, 2000 (1996).  
  
\bibitem{jancovici77}  
B.\ Jancovici J.\ Stat.\ Phys.\ {\bf 17}, 357 (1977).  
  
\bibitem{ds03}  
H.\ DeWitt and W.\ Slattery, Contrib.\ Plasma Phys.\ {\bf 43}, 279 (2003).  
  
\bibitem{pc00}  
A.~Y.\ Potekhin and G.~Chabrier, Phys.\ Rev.\ {\bf E62},  
8554 (2000).  
  
\bibitem{ds99}  
H.\ DeWitt and W.\ Slattery, Contrib.\ Plasma Phys.\ {\bf 39}, 97 (1999).  
  
\bibitem{oii91}  
S.\ Ogata, H.\ Iyetomi and S.\ Ichimaru, Astrophys.\ J.\  
{\bf 372}, 259 (1991)  
  
\bibitem{oiv93}  
S.~Ogata, S.~Ichimaru and H.~M.~Van Horn, Astrophys.\ J.\  
{\bf 417}, 265 (1993).  
  
\bibitem{st83}  
S.\ L.\ Shapiro and S.\ A.\ Teukolsky, {\em Black Holes, White Dwarfs,  
and Neutron Stars}, Wiley-Interscience, New York, 1983.  
  
\bibitem{baiko}  
D.~A.\ Baiko, Phys.\ Rev.\ {\bf E66}, 056405 (2002).  
  
\bibitem{iovh92}  
S.~Ichimaru, S.~Ogata and H.~M.~Van Horn, Astrophys.\ J.\ {\bf 401},  
L35 (1992).  
  
\bibitem{pm04}  
E.~L.~Pollock and B.~Militzer,  
Phys.\ Rev.\ Lett.\ {\bf 92}, 021101 (2004).  
  
\bibitem{ki95}  
H.~Kitamura and S.~Ichimaru,  
Astrophys.\ J.\ {\bf 438}, 300 (1995).  
  
\bibitem{kitamura00}  
H.\ Kitamura, Astrophys.\ J.\ {\bf 539}, 888 (2000).  
  
\bibitem{baikoetal98}  
D.\ A.\ Baiko, A.\ D.\ Kaminker, A.\ Y.\ Potekhin and  
D.\ G.\ Yakovlev, Phys.\ Rev.\ Lett.\ {\bf 81}, 5556 (1998).  
  
\bibitem{kaminkeretal99}  
A.\ D.\ Kaminker, C.\ J.\ Pethick, A.\ Y.\ Potekhin,  
V.\ Thorsson and D.\ G.\ Yakovlev, A\&A {\bf 343},  1009 (1999).  
  
\bibitem{ik99}  
S.~Ichimaru and H.~Kitamura,  
Phys.\ Plasmas {\bf 6}, 2649 (1999); erratum: {\bf 7}, 1335 (2000).  
  
\bibitem{filippenko04}  
A.\ V.\ Filippenko, in: White Dwarfs: Probes of Galactic  
Structure and Cosmology, eds.\ E.M.\ Sion, H.L.\  
Shipman and S.\ Vennes (Kluwer: Dordrecht) 2004,  
in press [astro-ph/0410609].  
  
\bibitem{itohetal92}  
N.\ Itoh, H.\ Mutoh, A.\ Hikita and Y.\ Kohyama,  
Astrophys.\ J.\ {\bf 395}, 622 (1992);  
erratum: {\bf 404}, 418 (1993).  
  
\bibitem{ry82}  
M.\ E.\ Raikh and D.\ G.\ Yakovlev, Astrophys.\ Space Sci.\  
{\bf 87}, 193 (1982).  
  
\bibitem{woos04} S.\ E.\ Woosley, A.\ Heger, A.\ Cumming, R.\ D.\ Hoffman, 
J.\ Pruet, T.\ Rauscher, J.\ L.\ Fisker, H Schatz, B.\ A.\ Brown, and M.\
Wiescher, Astrophys.\ J.\ Suppl.\ {\bf 151}, 75 (2004). 
  
\end{thebibliography}
\end{document}